\documentclass[reprint,amsmath,amssymb,aps]{revtex4-2}

\usepackage{bm}
\usepackage{graphicx}
\graphicspath{{plots/}{../plots/}}
\usepackage{subfigure}

\usepackage[ruled,vlined]{algorithm2e}

\SetKwInput{KwIn}{Input}
\SetKwInput{KwOut}{Output}
\SetKwInput{KwInit}{Init}

\begin{document}

\title{Renormalization group for spectral collapse in random matrices\\with power-law variance profiles}
\author{Philipp Fleig}
\email{philipp.fleig@mr.mpg.de}
\affiliation{Department of Cellular Biophysics, Max Planck Institute for Medical Research,\\ Jahnstra{\ss}e 29, 69120 Heidelberg, Germany}

\date{\today}

\begin{abstract}
We propose a renormalization group (RG) approach to compare and collapse eigenvalue densities of random matrix models of complex systems across different system sizes. The approach is to fix a natural spectral scale
by letting the model normalization run with size, turning raw spectra into comparable, collapsed density curves. We demonstrate this approach on generalizations of two classic random matrix ensembles---Wigner and Wishart---modified to have power-law variance profiles. 
We use random matrix theory methods to derive self-consistent fixed-point equations for the resolvent to compute their eigenvalue densities, we define an RG scheme based on matrix decimation, and compute the Beta function controlling the RG flow as a function of the variance profile power-law exponent.
The running normalization leads to spectral collapse which we confirm in simulations and solutions of the fixed-point equations.
We expect this RG approach to carry over to other ensembles, providing a method for data analysis of a broad range of complex systems.
\end{abstract}

\maketitle

\section{Introduction}

Many large data sets of complex systems with $N$ interacting units 
are naturally summarized by matrices: correlations between neurons~\cite{CunninghamYu2014} or genes~\cite{Eisen1998}, couplings in materials~\cite{AshcroftMermin1976}, network connectivities~\cite{AlbertBarabasi2002}, or interaction kernels in collective behaviour~\cite{Bialek2012Flocks}. A compact, system‑level descriptor is the spectral distribution—the distribution of matrix eigenvalues—which encodes properties such as variability, stability, and characteristic scales of the system~\cite{Wigner_1958,Marchenko_Pastur1967,Sompolinsky1988,mastrogiuseppe2018linking}. In practice, across different studies or different runs of an experiment, data matrices may be recorded or simulated at different system sizes $N$. However, one would like to extract properties of the system in a size-independent way. This raises a basic technical question: how should one compare spectra across matrix sizes, and how can they be \emph{collapsed} onto a common reference curve?

Random matrix theory (RMT) provides a useful language to address this question. It offers a foundation for high‑dimensional statistics with macroscopic laws that are universal, i.e. insensitive to microscopic details~\cite{Wigner_1958,BaiSilverstein_book,AGZ2010,potters2020first,couillet2011random}. Predictions include principled null models, signal–noise separation, detection and stability thresholds, and phase transitions in high‑dimensional inference~\cite{Marchenko_Pastur1967,Laloux_etal_1999,Fleig_Nemenman_2022,Sompolinsky1988,zdeborova2016statistical,Lesieur2017}. Random matrix theory also supplies standard normalizations and finite‑$N$ edge fluctuations—e.g. Wigner’s variance scaling, the aspect ratio in Mar\v{c}enko--Pastur, and soft Tracy--Widom edges~\cite{Wigner_1958,Marchenko_Pastur1967,TracyWidom1994,potters2020first}.
However, these asymptotic normalizations and edge scalings standardize particular regimes (e.g. bulk and microscopic edge regimes), but they do not provide a general scale‑setting rule for collapsing entire spectral curves across different matrix sizes.

We address this gap by adopting a \emph{renormalization group} (RG) viewpoint~\cite{PhysicsPhysiqueFizika.2.263,Kogut1979RG_Epsilon,goldenfeld2018lectures}. The idea is simple and operational: choose a natural spectral scale (e.g. mean level, bandwidth, etc.) to hold fixed as the matrix size changes—analogous to a renormalization condition—and let a single normalization factor run with matrix size under a coarse‑graining scheme. We show that with such a running normalization, spectra at different sizes collapse and can be compared on equal footing, while the residual size‑dependence is funneled into a single running coupling governed by a Beta function~\cite{WilsonKogut1974}. We note that an RG‑flavored perspective of RMT in high‑dimensional regression has been explored in~\cite{atanasov2024scaling}.

We illustrate the recipe on two generalizations of classic random matrix (RM) ensembles—Wigner~\cite{Wigner_1958} and Wishart~\cite{Marchenko_Pastur1967}—
with \emph{power‑law variance profiles}. Ensembles with such a power-law structure are motivated by heavy‑tailed participation and coupling strengths in neural circuits and networks~\cite{buzsaki2014log,RajanAbbott2006,MasudaPorterLambiotte2017}, and by heterogeneous sampling in high‑dimensional data~\cite{Alter2000,CunninghamYu2014,luecken2019current}.
Using RMT methods, we derive self‑consistent fixed-point equations for the finite‑$N$ RM resolvent, solve them numerically to obtain the eigenvalue density, and find excellent agreement with simulations. We then derive the running normalization that fixes a chosen spectral scale, demonstrate spectral collapse, define a coarse-graining scheme (via matrix decimation) and compute the Beta function to delineate relevant, marginal, and irrelevant regimes of the RG flow.
At the marginal exponent, the Wigner-type ensemble has a non-trivial RG fixed point, whereas the Wishart-type ensemble exhibits marginal flow with logarithmic corrections and approaches the trivial limit only logarithmically slowly.
Finally, we discuss the large‑$N$ limit where the coupling flow is governed by a Callan–Symanzik equation~\cite{Callan1970CS,Symanzik1970CS} and derive an integral equation for the large-$N$ RM resolvent. We will provide full details for the Wigner-type ensemble and present the key results for the Wishart-type ensemble.

\section{Wigner ensemble with power-law variance profile}

\subsection{Ensemble definition}
Given system size $N\in\mathbb{N}$, we construct the symmetric $N\times N$ random matrix $\mathbf X_N=(X_{jk})$ with independent centered entries:
\begin{align}
    \mathbb{E}[X_{jk}]&=0\,,\quad X_{jk}=X_{kj}\,,\quad 1\le j\le k\le N\,,\label{eq:model_eq1}\\
    \mathrm{Var}(X_{jk})&=\frac{\sigma^2}{\gamma}\,j^{-\alpha}k^{-\alpha}\,,\quad j<k\,,\label{eq:model_eq2}\\
    \mathrm{Var}(X_{jj})&=\frac{2\sigma^2}{\gamma}\,j^{-2\alpha}\label{eq:model_eq3}\,,
\end{align}
where $\alpha\in\mathbb{R}_{\geq0}$ is the power-law exponent, $\sigma^2>0$ a constant variance scaling factor, and $\gamma>0$ the normalization factor whose form we are going to determine through our RG analysis. Thus, the RM ensemble consists of matrices with independent, centered entries and variances as defined in Eqs.~\eqref{eq:model_eq2}–\eqref{eq:model_eq3}. The variances follow a power-law profile that decays as we move diagonally across the matrix, starting from the upper left corner.
For $\alpha=0$ the variance profile becomes constant across matrix entries and by setting the normalization to scale linearly with matrix size, $\gamma\equiv N$, the definition of the standard Wigner ensemble is recovered~\cite{Wigner_1958,potters2020first}.

\subsection{RMT definitions}
Our first goal will be to determine the eigenvalue density and we start by setting some notation. For a random matrix $\mathbf X_N$ of size $N\times N$, we define the \emph{empirical resolvent} (or Green's function) as
\begin{align}\label{eq:resolvent}
    \mathbf G_N^{\mathrm{emp}}(z):=\big(z\,\mathbf I_N-\mathbf X_N\big)^{-1}\,,\quad z=x+i\eta\,,\ \eta>0\,.
\end{align}
Taking the normalized trace yields at finite $N$:
\begin{align}\label{eq:rand_Stieltjes_transform}
    G_N^{\mathrm{emp}}(z):=\frac1N\,\mathrm{Tr}\,\mathbf G_N^{\mathrm{emp}}(z)\,.
\end{align}
For a self-averaging normalized trace, $G_N^{\mathrm{emp}}(z)$ concentrates around its \emph{deterministic equivalent}, $G_N(z)$
\begin{align}
G_N^{\mathrm{emp}}(z)\simeq G_N(z)\,,
\end{align}
up to fluctuations of order $(N\eta)^{-1/2}$ (see Appendix~\ref{app:local_laws}). The finite-$N$ eigenvalue density is obtained as
\begin{align}\label{eq:rho_N}
    \rho_{N,\eta}(x)=-\,\frac1\pi \,\Im\, G_N(x+i\,\eta)\,,
\end{align}
and analogously from $G_N^{\mathrm{emp}}$, up to errors. If $G_N(z)$ converges in the large-$N$ limit, we denote the limit by $\overline{G}(z)$ and obtain the eigenvalue density from $\rho(x)=-\,\tfrac{1}{\pi}\,\Im\,\overline{G}(x+i0^+)$. In what follows we work with the deterministic fixed-point objects and only write out the label ``$\mathrm{emp}$" explicitly, when needed to avoid ambiguity.

In the derivation of deterministic fixed-point equations for the resolvent, we will for simplicity sometimes assume that the matrix entries $X_{ik}$ are Gaussian. However, we note that the fixed‑point structure depends only on the variance profile rather than exact Gaussianity. The same set of equations can be derived without assuming Gaussianity (assuming standard moment and tail control~\cite{ajanki2017universality,ajanki2019stability,KnowlesYin2017Anisotropic} instead). In simulations we use Gaussian entries. Additionally, we provide numerical evidence hinting at insensitivity of the spectral density to the entry law within a finite‑variance class (including a heavy‑tailed example). Throughout the article we adopt physics-style reasoning, instead of full mathematical rigor.

\subsection{Fixed-point equation}
Our first goal is to write down a system of \emph{self-consistent} equations for the entries of the resolvent and solve it to obtain the eigenvalue density. Here, we provide a heuristic derivation of the self-consistent equations, giving the full-length derivation in Appendix~\ref{app:Wigner-type-eqn}. 
We start from the \emph{matrix Dyson equation} (MDE)~\cite{ajanki2019stability,alt2020dyson} for the deterministic equivalent:
\begin{align}
\mathbf G_N(z)^{-1} &= z\,\mathbf I_N - \boldsymbol\Sigma(z)\,,\text{ with }\boldsymbol\Sigma(z) = \mathcal S\!\big[\mathbf G_N(z)\big]\,,
\end{align}
where $\boldsymbol\Sigma$ is the Dyson self–energy, and is given by a variance operator $\mathcal S$ (determined by second moments) acting on the resolvent.

To treat the Wigner ensemble with power-law variance profile it is convenient to introduce the following notation
\begin{align}
    w_j&:=j^{-\alpha}\,,\quad w:=(w_1,\ldots,w_N)^T\,,\\
    \mathbf D&:=\mathrm{diag}(w_1^2,\ldots,w_N^2)\,,\label{eq:a_D}
\end{align}
where we refer to the $w_j$'s as variance weights. The variance operator $\mathcal S$ for the Wigner ensemble with power-law profile is given by the variance matrix $\mathbf S$ which, by Eqs.~\eqref{eq:model_eq2}-\eqref{eq:model_eq3}, is given by
\begin{align}\label{eq:var_matrix}
    s_{jk}:=\begin{cases}
    \frac{\sigma^2}{\gamma}\,w_j w_k\,,& j\neq  k\\[6pt]
    \frac{2\sigma^2}{\gamma}\,w_j^2\,,& j=k
    \end{cases}
    \quad \Leftrightarrow\quad \mathbf S:=\frac{\sigma^2}{\gamma}[w w^T + \,\mathbf D]\,.
\end{align}
This variance operator preserves diagonality.
In other words, at the deterministic fixed point, both $\mathbf G_N$ and $\boldsymbol\Sigma=\mathbf S\mathbf G_N$ are diagonal
\begin{align}
    \big(\mathbf S\mathbf G_N\big)_{jj}&=\sum_{k=1}^N s_{jk}\,(\mathbf G_N)_{kk}(z)\,,\quad
    \big(\mathbf S\mathbf G_N\big)_{ij}=0\,,
\end{align} 
for $i\neq j$ and the self-consistent equations close on the diagonal
\begin{align}
    (\mathbf G_N)_{jj}(z)=\frac{1}{z-\sum_{k=1}^N s_{jk}\,(\mathbf G_N)_{kk}(z)}\,.
\end{align}
With $s_{ik}$ from Eq.~\eqref{eq:var_matrix} this becomes
\begin{align}
\label{eq:Dyson_equation_rank1_diag}
    \frac{1}{G_j}=z- \frac{\sigma^2}{\gamma}\big[w_j \sum_{k=1}^N w_k G_k+\,w_j^2\,G_j\big]\,,
\end{align}
where we have defined, $G_j:=(\mathbf G_N)_{jj}$.
From this equation, we define two parameters
\begin{align}
    \label{eq:order_param}
    c:=\frac{\sigma^2}{\gamma}\,,\quad\text{and }\quad u(z):=\sum_{k=1}^N w_k\,G_k(z)\,,
\end{align}
which we refer to as the \emph{coupling} and the \emph{variance‑weighted mean field}, respectively.
In Eq.~\eqref{eq:Dyson_equation_rank1_diag}, site $j$ experiences the variance‑weighted mean field, $w_j\,u(z)$, while the diagonal contribution, $w_j^2\,G_j$, is a local term.
Overall, for fixed $u(z)$, each $G_j$ solves a scalar quadratic equation
\begin{align}
\label{eq:quad_Gj}
    \frac{1}{G_j}=z- c(w_j u + w_j^2G_j)\,,
\end{align}
with the physical branch given by $\Im\,G_j(z)<0$.
Equations~\eqref{eq:order_param}--\eqref{eq:quad_Gj} reduce the problem to a one-dimensional, self-consistent fixed-point equation for $u(z)$, from which $G_j(z)$ and hence also the normalized trace, $G_N(z)$, follow. The fixed point serves as a deterministic predictor that approximates $G^\mathrm{emp}_N(z)$ up to local-law errors, cf. Appendix~\ref{app:local_laws}.

As a sanity check, we verify the reduction to the Wigner semicircle law.
For $\alpha=0$, $w_j= 1$ for all $j$. By symmetry, $G_j(z)= G_N(z)$, and $u= N G_N$. Summing the self-consistent equation in Eq.~\eqref{eq:quad_Gj} over sites, we obtain a scalar Dyson equation of the form
\begin{align}
    \frac{1}{G_N}= z- c(u+ G_N)= z- c\,(N+1)G_N\,,
\end{align}
such that
\begin{align}
\label{eq:quad_wigner}
    G_N(z)=\frac{z\,\pm\,\sqrt{z^2-R^2}}{2\,c\,(N{+}1)}\,,\quad R=2\,\sqrt{\frac{\sigma^2}{\gamma}(N{+}1)}\,. 
\end{align}
Thus, the limiting eigenvalue density is the semicircle with radius $R$. Keeping the normalization $\gamma$ general makes the scaling explicit. The standard Wigner scaling $\gamma=N$ yields $R\to 2\sigma\,\sqrt{1+1/N}\to 2\sigma$ as $N\to\infty$, i.e. the classical Wigner semicircle law with variance parameter $\sigma^2$.

\begin{figure}[t]
\centering
\subfigure[Linear-linear scale.]{
\label{fig:panel}
\includegraphics[width=8.4cm]{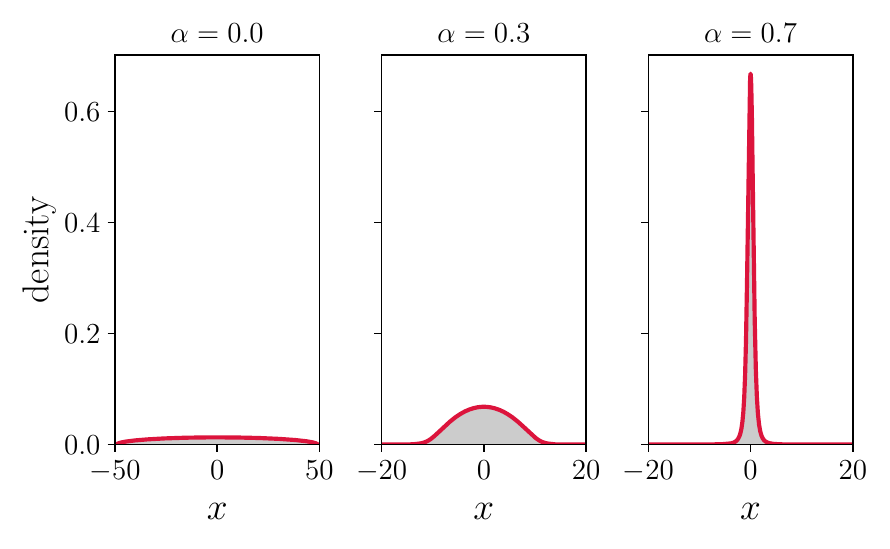}
}
\\
\subfigure[Log-linear scale.]{
\label{fig:panel_loglog}
\includegraphics[width=8.4cm]{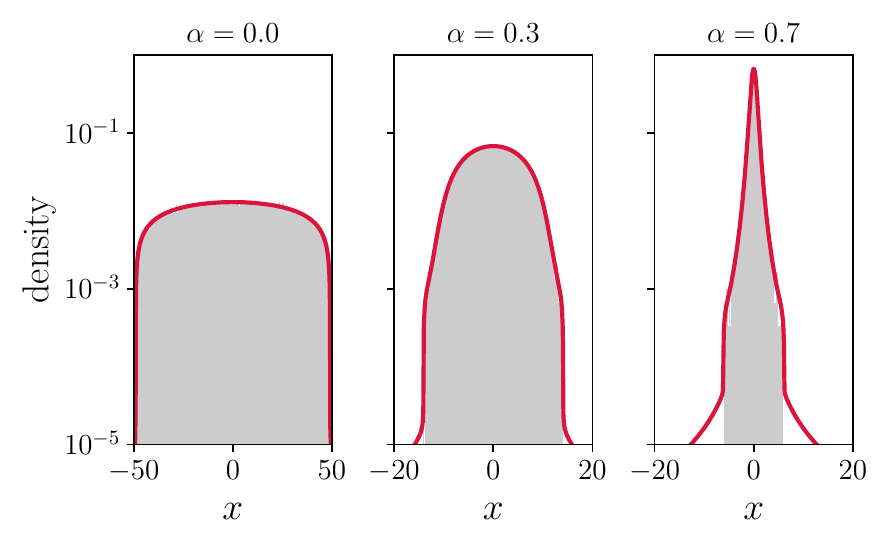}
}
\caption{{\bf Eigenvalue densities of Wigner ensemble with power-law variance profile.} (a) and (b) Eigenvalue density for different values of $\alpha$ from simulations (gray) and fixed-point solutions of the self-consistent finite-$N$ equations (red). Across all plots we use $\gamma=1$, $N=600$, $\sigma^2=1$, $20$ independent simulation trials, and $\alpha$ values as indicated on the subplots. All curves are Cauchy-smoothed with kernel width $\eta=10^{-2}$. Linear-linear scale (a) and log-linear scale (b).}
\end{figure}
An algorithm for solving the fixed-point equations~\eqref{eq:order_param}--\eqref{eq:quad_Gj} by fixed-point iteration is described in Algorithm~\ref{algo:algorithm_1} of Appendix~\ref{app:algorithm-wigner}. In Figs.~\ref{fig:panel} and~\ref{fig:panel_loglog} we compare the eigenvalue density $\rho_{N,\eta}$ computed from the solution of the fixed-point equations to eigenvalue histograms obtained from simulations with the RM $\mathbf X_N$ simulated across multiple independent trials.  We find a near-perfect match between fixed-point solutions (red curves) and simulations (gray) for exponent values $\alpha=0.0$, $0.3$, and $0.7$. The differences between fixed-point solutions and simulation data at the edges are an artifact due to leakage of the Cauchy smoothing (see Appendix~\ref{app:visualization}). Figures~\ref{fig:panel} and~\ref{fig:panel_loglog} show the \emph{raw} eigenvalue densities with trivial normalization factor $\gamma=1$. We note that the support of these raw densities changes significantly across the different values of $\alpha$. 

\section{Renormalization Group analysis}
\begin{figure*}[t]
\centering
\subfigure[Raw and collapsed densities (linear-linear scale).]{
\label{fig:collapse-beta_lin}
\includegraphics[width=8.cm]{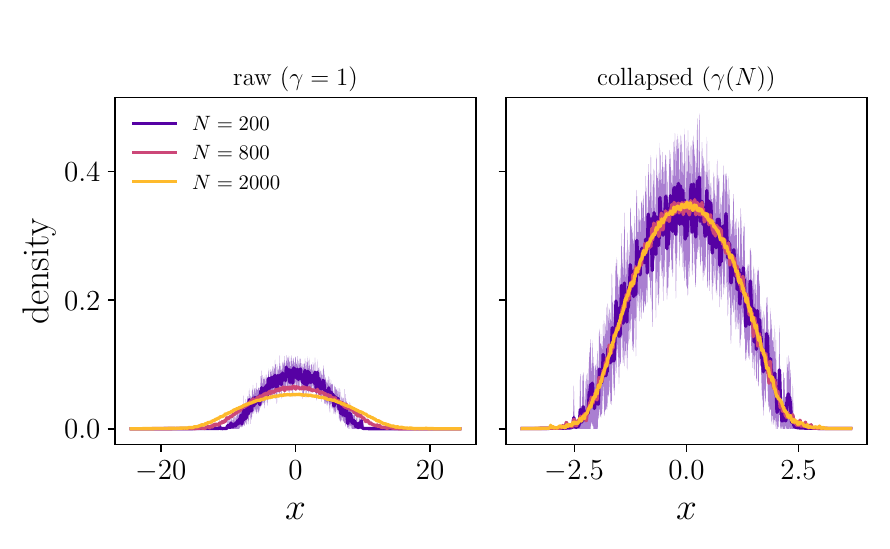}
}
\hspace{1cm}
\subfigure[Raw and collapsed densities (log-linear scale).]{
\label{fig:collapse-beta_log}
\includegraphics[width=8.cm]{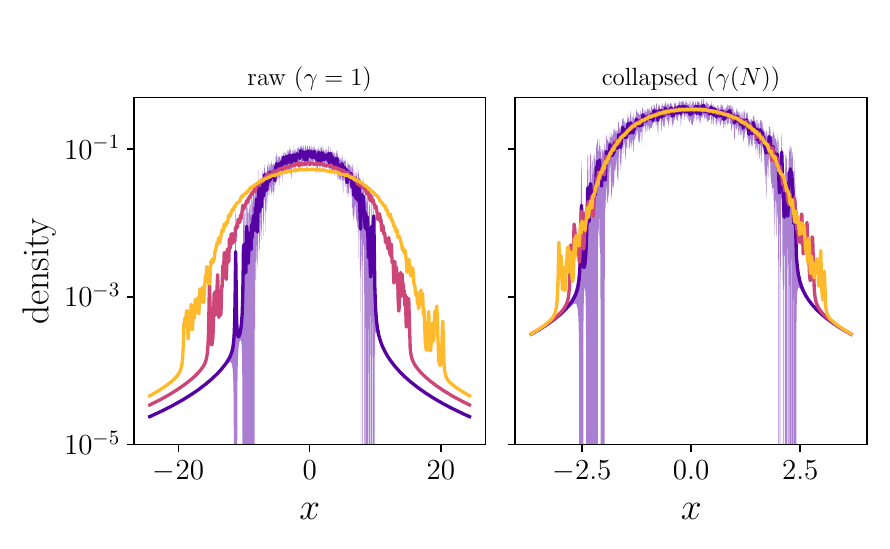}
}\\

\subfigure[Normalization flow with decimation.]{
\label{fig:gamma_flow}
\includegraphics[width=7cm]{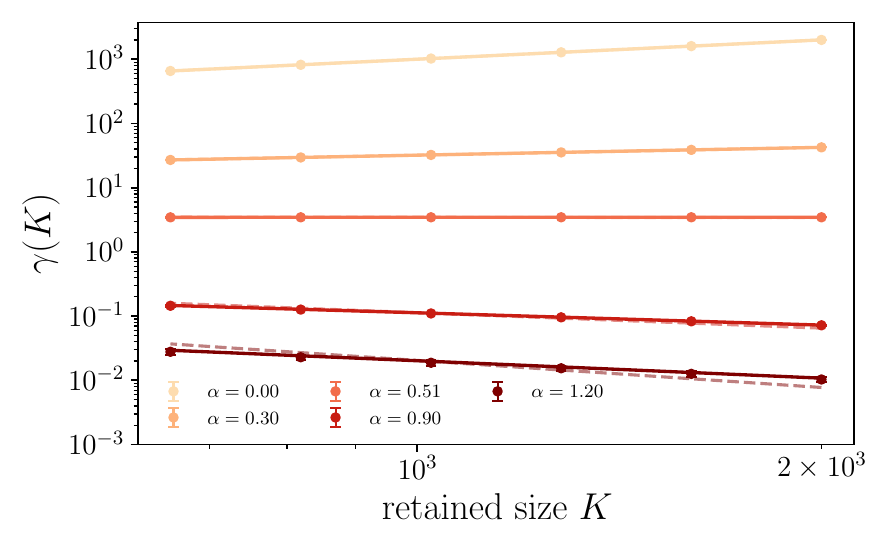}
}
\hspace{2cm}
\subfigure[Coupling flow with decimation.]{
\label{fig:decimation_flow}
\includegraphics[width=7cm]{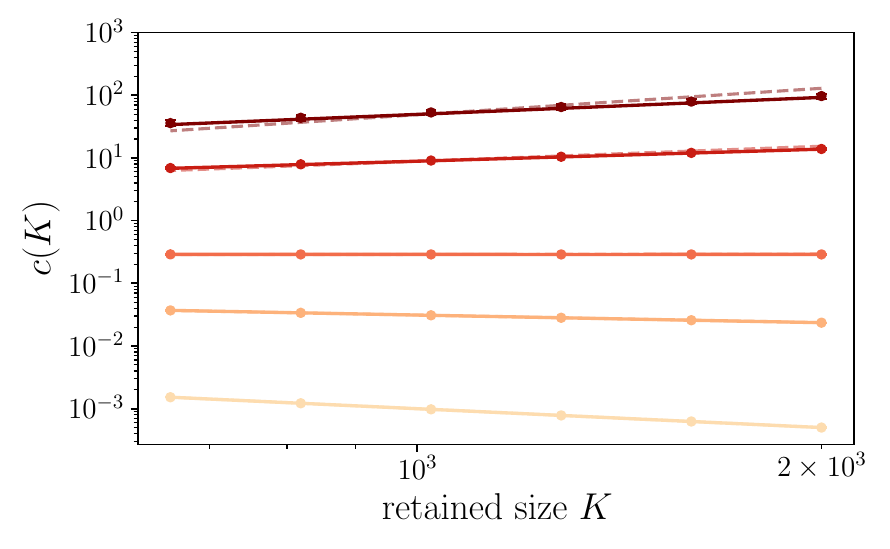}
}
\caption{{\bf Eigenvalue density collapse and running normalization.}  (a) and (b) Raw and collapsed eigenvalue densities obtained from simulations of the ensemble. The raw densities (left subplot) use $\gamma=1$ and the collapsed densities are obtained using the running normalization $\gamma(N)$. We use $\alpha=0.3$ and $N=200$, $800$, and $2000$ and ten independent ensemble trials. For $N=200$ we also show the one standard deviation band of sampling noise. Linear-linear scale (a) and log-linear scale (b). (c) Decimation flow of normalization: $\gamma(K)$ versus retained size $K$ after removing a fixed fraction of largest indices at each step, while the second moment is kept fixed at unity. Solid lines: exact prediction. Markers with error bars: simulation with mean and standard deviation from six independent trials. Dashed lines serve as guides indicating slope $1-2\alpha$. Dashed lines are mostly covered by the exact prediction. (d) Decimation flow of the coupling $c$.}
\end{figure*}
We now turn to an RG analysis of the power-law-structured Wigner ensemble and start by deriving a running normalization $\gamma(N)$ that collapses eigenvalue densities onto each other. 
\subsection{RG condition and spectral collapse}
The second moment of $\mathbf X_N$ is given by
\begin{align}\label{eq:second_moment_X}
    m_2=\frac{1}{N}\,\mathbb{E}[\mathrm{Tr}\,\mathbf X_N^2]\,,
\end{align}
which equals the expected average of squared eigenvalues (the first moment vanishes by Eq.~\eqref{eq:model_eq1}). We choose the second moment (the typical bandwidth) as the spectral scale that we want to keep constant under renormalization. Fixing the second moment to a constant value $m_2^\star$:
\begin{align}
\label{eq:rg_condition}
    m_2 \stackrel{!}{=} m_2^\star\,,
\end{align}
is analogous to fixing an energy density per degree of freedom in statistical mechanics or a variance per site in disordered systems~\cite{Kogut1979RG_Epsilon,MezardParisiVirasoro1987}. In the next section we introduce an RG scheme and we are going to interpret Eq.~\eqref{eq:rg_condition} as a {\it renormalization condition}. To satisfy Eq.~\eqref{eq:rg_condition} we introduce the normalization $\gamma(N)$ which \emph{runs} with system size and equally the running coupling, $c(N)=\sigma^2/\gamma(N)$, cf. Eq.~\eqref{eq:order_param}. It is natural to impose $m_2^\star$ to be of order unity and without loss of generalization, we set $m_2^\star\equiv 1$ in all calculations and figures.

Using the ensemble symmetry $X_{jk}=X_{kj}$, from Eq.~\eqref{eq:second_moment_X} we compute the following expression for the second moment
\begin{align}\label{eq:second_moment}
m_2=c\,r_N(\alpha)\,,
\end{align}
where
\begin{align}
  r_N(\alpha)&:=\frac{1}{N}\Big[\big(\sum_{j=1}^N w_j\big)^2 + \sum_{j=1}^N w_j^2\Big]\nonumber\\
  &=\frac{H_N(\alpha)^2+H_N(2\alpha)}{N}\,,
\end{align}
and we have used the definition of the {\it generalized harmonic sum}
\begin{align}
    H_N(\theta):=\,\sum_{j=1}^N j^{-\theta}\,.
\end{align}
With the definition of the coupling in Eq.~\eqref{eq:order_param}, and the renormalization condition Eq.~\eqref{eq:rg_condition}, Eq.~\eqref{eq:second_moment} implies:
\begin{align}\label{eq:gamma_N}
  \gamma(N)&=\frac{\sigma^2}{m_2^\star}\,r_N(\alpha)\,,
\end{align}
such that the normalization is $N$-dependent.
For the RG analysis we are interested in the dominant scaling behaviour with $N$. We note the following scaling regimes of the generalized harmonic sum: $H_N(\theta)\sim \tfrac{N^{1-\theta}}{1-\theta}$ for $\theta<1$, $H_N(\theta)\sim \log N$ for $\theta=1$, and $H_N(\theta)\to \zeta(\theta)$ for $\theta>1$, cf. Appendix~\ref{app:large_N}. We find the dominant large-$N$ behaviour of $r_N(\alpha)$ is
\begin{align}
  r_N(\alpha)\sim
  \begin{cases}
    \dfrac{1}{(1-\alpha)^2}\,N^{1-2\alpha}, & 0\le\alpha<1,\ \alpha\neq1/2\,,\\[6pt]
    \mathrm{const.}, & \alpha=1/2\,,\\[6pt]
    \dfrac{\log^2 N}{N}, & \alpha=1\,,\\[6pt]
    N^{-1}, & \alpha>1\,,
  \end{cases}
\end{align}
which we confirm numerically in Appendix Fig.~\ref{fig:m2_scaling}.
Thus, we find $\alpha$-dependent power-law scaling of the normalization with the system size $N$ for $\alpha\geq 0$, except at $\alpha=1/2$ where $\gamma(N)=O(1)$, and a logarithmic correction at $\alpha=1$.
We note that by Eq.~\eqref{eq:order_param} the coupling scales inverse to the normalization.

With the normalization $\gamma(N)$ chosen to keep the second moment constant, the finite-$N$ spectral densities obtained from simulations of the ensemble at different values of $N$ and fixed exponent $\alpha$ \emph{collapse} onto a common curve. In Figs.~\ref{fig:collapse-beta_lin}--\ref{fig:collapse-beta_log} we show a comparison of raw and collapsed eigenvalue densities. For $N=200$, we also show the one standard deviation band due to simulation sampling noise. While the raw curves (left subplots with $\gamma=1$ normalization) generally do not lie within the sampling noise band, the collapsed curves (right subplots with $\gamma(N)$ normalization) do and largely collapse onto each other. From the collapsed curves we observe some small effects of non-uniform edge convergence. Overall, this demonstrates approximate spectral invariance along the RG trajectory in finite-$N$ simulations: after fixing the second moment via the running normalization $\gamma(N)$, the eigenvalue density is independent of system size up to finite‑$N$ corrections. We note that the collapse discussed above concerns the bulk density. For $\alpha>1$ the bulk density degenerates to a delta mass at zero, and only a finite number of $O(1)$--outliers associated with small index values remain. These outliers are not self-averaging.

\subsection{Decimation (coarse-graining) procedure}
Next, we introduce an RG coarse-graining procedure via matrix {\it decimation} and interpret the running normalization and coupling in this context. We define that one decimation step consists of removing a fraction of matrix indices corresponding to the smallest weights $w_j$ (largest index values). We fix a scale factor $b>1$ and retain indices $j\le \lfloor N/b\rfloor$. The retained block, $\mathbf X^{(\mathrm{dec})}$, inherits the same variance profile with weights $w^{(\mathrm{dec})}_j = j^{-\alpha}$ with $j=1,\ldots,\lfloor N/b\rfloor$, up to rescaling.
From Eq.~\eqref{eq:second_moment} we obtain the coarse-grained second moment
\begin{align}
    m_2'=\frac{1}{N/b}\,\mathbb{E}\,[\mathrm{Tr}\,{\mathbf X^{(\mathrm{dec})}}^2]= c'\,\frac{H_{N/b}(\alpha)^2 + H_{N/b}(2\alpha)}{N/b}\,,
\end{align}
where we implicitly apply the floor function to non-integer values of $N/b$.
To enforce the renormalization condition in Eq.~\eqref{eq:rg_condition} under one decimation step $N\mapsto N/b$, we choose the renormalized coupling $c'$ so that $m_2'(N/b;c')=m_2(N;c)$. This yields the discrete RG update for the running coupling and normalization:
\begin{align}
\label{eq:flow_c}
    \frac{c'}{c}=\frac{\gamma}{\gamma'} \sim \frac{1}{b}\,\frac{H_{N}(\alpha)^2 + H_{N}(2\alpha)}{H_{N/b}(\alpha)^2 + H_{N/b}(2\alpha)} \;\xrightarrow[N\to\infty]{}\; b^{1-2\alpha}\,.
\end{align}

If we take a total of $n$ decimation steps with fixed scale factor $b>1$, the retained system size $K_n$ is
\begin{align}\label{eq:K_n}
    K_n := \Big\lfloor \frac{N}{b^{n}}\Big\rfloor\quad \Leftrightarrow\quad n = \frac{\log(N/K_n)}{\log b}\,,
\end{align}
with $K_0=N$. Iterating the update in Eq.~\eqref{eq:flow_c} yields the $n$-step flow for the coupling
\begin{align}
  \frac{c_n}{c_0}
  \sim \prod_{\bar n=0}^{n-1} \frac{1}{b}\,
  \frac{H_{N/b^{\bar n}}(\alpha)^2 + H_{N/b^{\bar n}}(2\alpha)}{H_{N/b^{\bar n+1}}(\alpha)^2 + H_{N/b^{\bar n+1}}(2\alpha)}
  \;\xrightarrow[N\to\infty]{}\; b^{\,n(1-2\alpha)}\,,
\end{align}
such that asymptotically, $c_n \propto b^{n(1-2\alpha)}$, or, using Eq.~\eqref{eq:K_n}: 
\begin{align}
  c_n \propto \Big(\frac{N}{K_n}\Big)^{1-2\alpha}
  \quad\Rightarrow\quad
  \gamma(K_n;N)=\frac{\sigma^2}{c_n}\propto K_n^{\,1-2\alpha}\,,
\end{align}
where the proportionality is up to a constant that does not depend on $K_n$ (for fixed initial $K_0=N$).

In Figs.~\ref{fig:gamma_flow}--\ref{fig:decimation_flow}, we show the running of the normalization $\gamma(K)$ and coupling $c(K)$ with retained system size $K$, respectively.
Dashed lines serve as reference to indicate the theoretical scaling exponent, $1-2\alpha$, for the RG flow under decimation.
Observed deviations from the reference lines for $\alpha>1/2$ arise from finite-size effects, competition of scaling, and slow log-corrections.

\subsection{Beta function and RG regimes}
We define the running log-coupling, $\kappa:=\log c$, with $c(K_n)=\sigma^2/\gamma(K_n)$. Under one decimation step with per-step scale factor $b$, the discrete update is
\begin{align}
  \kappa_{n+1}-\kappa_n=(1-2\alpha)\,\log b
\end{align}
which integrates to
\begin{align}
  \kappa_n=\kappa_0+n(1-2\alpha)\,\log b\,.
\end{align}
Writing the total scale factor as, $B:=b^n$, and defining the RG scale, $s:=\log B$, the continuous infinitesimal flow along the RG trajectory with $m_2$ fixed becomes
\begin{align}\label{eq:beta_function}
  \beta^b_\mathrm{RG}(c):=\frac{d\kappa}{ds}=\frac{d\log c}{d\log B}=1-2\alpha\,.
\end{align}
This is the \emph{Beta function}~\cite{WilsonKogut1974} for $c$ of the RM ensemble.
It reveals three, physically distinct, regimes under decimation:

\emph{Relevant} ($\alpha<1/2$; $\beta^b_\mathrm{RG}>0$):
Under decimation to a smaller retained size $K$, $\gamma(K)\propto K^{1-2\alpha}$ decreases as $K$ decreases, while $c(K)$ increases under decimation. This is the RG sense in which the coupling is relevant.

\emph{Marginal} ($\alpha=1/2$; $\beta^b_\mathrm{RG}=0$): The coupling sits at an RG fixed point (a fixed line) and the Beta function changes sign at marginality. Since the fixed point remains interacting (non-vanishing self-energy and hence not the free resolvent, $1/z$), the fixed point is non-trivial. The normalization $\gamma(K)$ tends to a constant and the coupling $c(K)$ is scale-stationary. Any residual drift under decimation is a finite-size effect.

\emph{Irrelevant} ($\alpha>1/2$; $\beta^b_\mathrm{RG}<0$):
Under decimation to smaller $K$, $\gamma(K)$ increases and $c(K)$ decreases, i.e. the coupling weakens under decimation. As already noted above, for $\alpha>1$, the bulk density degenerates and non-self-averaging outliers can appear.

\subsection{RG flow and the Callan--Symanzik equation}\label{sec:cs}
\begin{figure}[t]
\centering
\includegraphics[width=8cm]{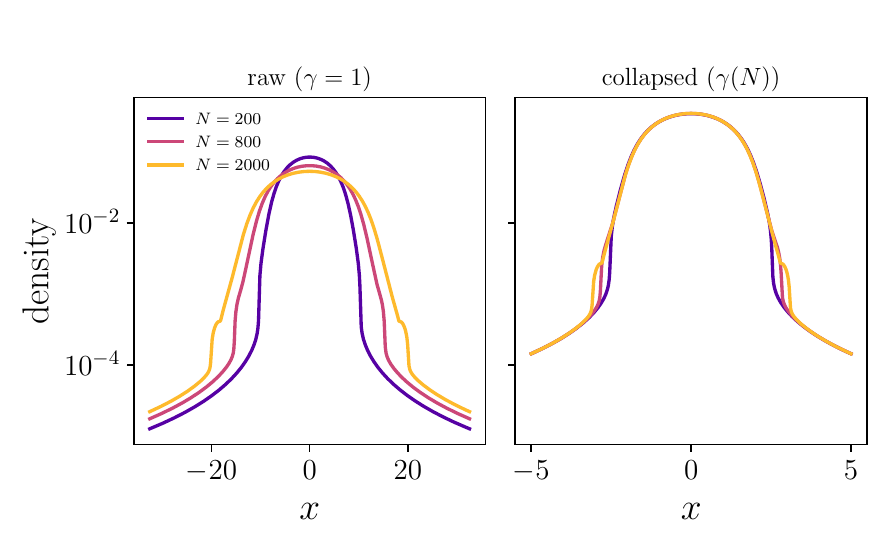}
\caption{{\bf Spectral collapse for the finite-$N$ self-consistent equation.} Spectral invariance test for $\alpha=0.30$ across $N=200$, $800$, and $2000$. Left: raw eigenvalue density $\rho_{N,\eta}(x)$ with $\gamma=1$ computed from the finite-$N$ self-consistent equation for different $N$. Right: collapsed eigenvalue density with running normalization $\gamma(N)$. We use $\eta_\mathrm{raw}=\sqrt{\gamma(N)}\eta_\mathrm{coll}$, with $\eta_\mathrm{coll}=10^{-2}$ for all $N$.
}\label{fig:cs_invariance}
\end{figure}
We now investigate the RG flow in the large-$N$ limit. 
For this, we set the RG scale, $s:=\log N$, and let the coupling run as $c(s)=\sigma^2/\gamma(s)$ so that the spectral scale (second moment) remains fixed along the flow. The running of the coupling is characterized by the Beta function
\begin{align}\label{eq:betaRG_s}
  \beta_{\mathrm{RG}} &:= \frac{d\log c}{d\log N}= -(1-2\alpha)
\end{align}
with a logarithmic correction at $\alpha=1$. In this alternative definition, the Beta function differs by a sign from the Beta function  in Eq.~\eqref{eq:beta_function}. Both are valid forms of the Beta function for the RM ensemble. 
Along this RG trajectory, spectral observables should be invariant to changes in the scale $s$ when the induced change in the coupling $c(s)$ is taken into account. One such observable is the mean resolvent, $G_N(z)=\frac{1}{N}\sum_{i=1}^N G_i(z)$. In the large-$N$ limit, we denote the mean resolvent by $\overline{G}$. In Appendix~\ref{app:large-N-limit-Wigner}, we derive an integral equation for $\overline G$ and solve it numerically.

Invariance along the flow is stated as, $d\overline{G}/ds=0$. Separating explicit and implicit scale dependence, through $c=c(s)$, yields the equation
\begin{align}
\label{eq:CS_equation}
  \big(\partial_s + \beta_{\mathrm{RG}}\,c\,\partial_c\big)\,\overline{G}(z;c) &= 0\,,
\end{align}
where we have used the Beta function in Eq.~\eqref{eq:betaRG_s}. This equation is known as the Callan--Symanzik equation~\cite{Callan1970CS,Symanzik1970CS} and holds in the large-$N$ limit along the RG trajectory. The equation captures how a shift in the scale $s$ is compensated by an induced change of the coupling $c$, leaving the mean resolvent $\overline{G}$ invariant.

At finite $N$, the quantity $d\,G_N/ds$,
is generally non-zero due to finite-size effects,
and the Callan--Symanzik equation is not satisfied exactly.
However, as we previously showed for simulated data in Figs.~\ref{fig:collapse-beta_lin}--\ref{fig:collapse-beta_log}, finite-size effects are controlled, implying that approximate invariance, and therefore collapse, of the eigenvalue density across different sizes $N$, is still observed for a running coupling.
As an additional step, we demonstrate the approximate invariance on the level of the resolvent.
For fixed $\alpha$, we solve the self-consistent Eqs.~\eqref{eq:order_param}--\eqref{eq:quad_Gj}, for different $N$ and the running normalization of Eq.~\eqref{eq:second_moment} and extract the density $\rho_{N,\eta}(x)$. In Fig.~\ref{fig:cs_invariance}, we show a comparison of raw and collapsed eigenvalue densities for three values of $N$. Up to non-uniform edge convergence, the curves collapse onto each other, confirming spectral invariance also for finite values of $N$ to good approximation. In Appendix~\ref{app:visualization}, we discuss our choice of $\eta$ values for the raw and collapsed curves.

The predictive power of the RG comes from \emph{universality}, which means that the RG flow is not an artifact of distributional choices of the model. For our RM ensemble this could mean that its universality class is determined by the power-law exponent and the multiplicative variance weights, but that spectral observables are largely insensitive to the particular choice of matrix entry law (subject to some conditions). In Appendix~Fig.~\ref{fig:universality_demos} we show a test of several entry-law distributions under the RG normalization that fixes the second moment and observe collapse across sizes in agreement with the solution of the self-consistent equation: Gaussian, Rademacher ($\pm1$), and Student-$t$ with finite variance (e.g. $\nu=4$) and the matrix entries are scaled by the variance weights $w_j$ to obtain the power-law variance profile. Within this ensemble class, the bulk of the eigenvalue density and edges appear insensitive to the entry-law after scale-fixing.

\section{Wishart ensemble with power-law variance profile}\label{subsec:wishart_brief}
\begin{figure*}[t]
\centering
\subfigure[Comparison between spectral distribution simulation (gray) and the self-consistent equation solution (red).]{
\label{fig:eval_Wishart}
\includegraphics[width=16cm]{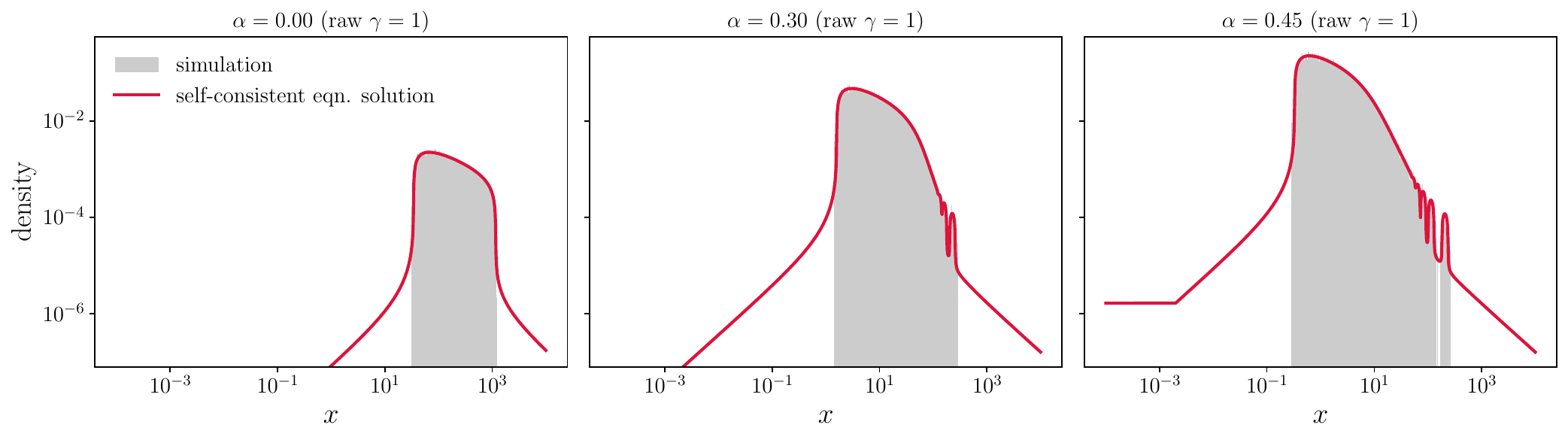}
}\\

\subfigure[Simulation spectral collapse.]{
\label{fig:wishart_simulation_collapse}
\includegraphics[width=8.cm]{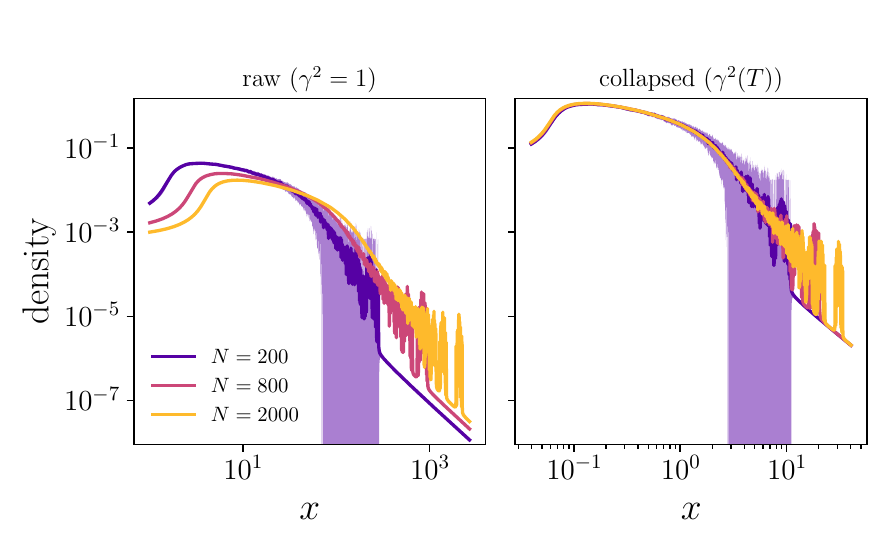}
}
\hspace{1cm}
\subfigure[Self-consistent equation spectral collapse.]{
\label{fig:wishart_de_collapse}
\includegraphics[width=8.cm]{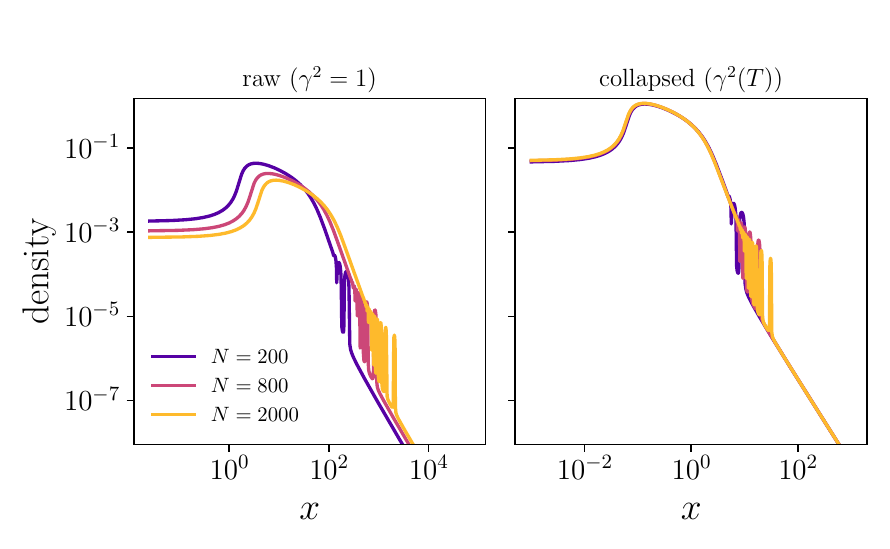}
}
\caption{{\bf Wishart ensemble with power-law variance profile.} (a) Comparison of spectral distribution simulations (gray histograms) and the fixed-point solution of the self-consistent equations (red). Across all subplots we use $\gamma=1$, $N=200$, $T=400$ (i.e. $q=0.5$), $\sigma^2=1$, and $\eta=10^{-2}$,  with $\alpha$ values as indicated above the plots. (b) Raw and collapsed eigenvalue densities obtained from simulations of the ensemble. We use $\alpha=0.3$ and $N=200$, $800$, and $2000$, and $q=0.5$ and simulate $100$ independent trials of the ensemble. (c) Raw and collapsed eigenvalue density from the fixed-point solution for the same parameters as (b). Left: raw densities with $\eta_\mathrm{raw}=\gamma^2(T)\, \eta_\mathrm{coll}$. Right: collapsed densities with $\eta_\mathrm{coll}=10^{-2}$ for all $N$.}
\end{figure*}

We study a Wishart ensemble with power-law heterogeneity placed in the \emph{rows} (e.g. observation index). Let $\mathbf Z\in\mathbb R^{T\times N}$ have i.i.d. centered unit-variance entries, let $\mathbf D:=\mathrm{diag}(v_1,\dots,v_T)$ where $v_t=t^{-2\alpha}$ with $t=1,2,\ldots,T$, and define
\begin{align}
  \mathbf Y := \mathbf D^{1/2}\mathbf Z\,,\quad
  \mathbf C := \frac{1}{\gamma^2}\,\mathbf Y^T \mathbf Y = \frac{1}{\gamma^2}\,\mathbf Z^T \mathbf D\,\mathbf Z \in\mathbb R^{N\times N}\,.
\end{align}
We also define the aspect ratio of matrix dimensions, $q:=N/T$ and think of $\mathbf Y$ as a $T\times N$ data matrix with variance decaying as a power-law across its rows, and $\mathbf C$ is the associated sample covariance matrix scaled by $\gamma^2$. In Appendix~\ref{app:wishart_finiteN_cavity} we give a cavity method derivation of the following finite‑$N$ self-consistent equation for the average resolvent of $\mathbf C$:
\begin{align}\label{eq:wishart_finiteN_G}
  G_N(z) &= \frac{1}{z - \Sigma_N(z)}\,,\\
  \Sigma_N(z) &= \frac{1}{\gamma^2}\sum_{t=1}^T \frac{v_t}{1 - (N/\gamma^2)\,v_t\,G_N(z)}\,.
  \label{eq:wishart_finiteN_Sigma}
\end{align}
For background on the derivation see~\cite{BaiSilverstein_book,SilversteinBai1995,DozierSilverstein2007,KnowlesYin2017Anisotropic,ajanki2019stability}. As a sanity check we note that for $\alpha=0$ the variance profile of $\mathbf Y$ becomes homogeneous and setting $\gamma^2 \equiv T$, the self-consistent equation reduces to a form that yields the well-known Mar\v{c}enko--Pastur distribution with aspect ratio $q$~\cite{Marchenko_Pastur1967,potters2020first}. An algorithm to solve the self-consistent equations~\eqref{eq:wishart_finiteN_G}-\eqref{eq:wishart_finiteN_Sigma} is described in  Algorithm~\ref{algo:algorithm_2} of Appendix~\ref{app:wishart_finiteN_cavity}. In Fig.~\ref{fig:eval_Wishart}, we show a comparison of the simulated eigenvalue density and the fixed-point solution obtained from the self-consistent equations and find an exact match. The differences between fixed-point solutions and simulation data at the edges are an artifact due to leakage of the Cauchy smoothing. The fixed-point solution (red curve) corresponds to the trial-averaged Cauchy-smoothed density at resolution $\eta$. Accordingly, the gray histogram aggregates eigenvalues across many independent realizations to approximate the same smoothed observable. In Appendix~\ref{app:large-N-limit-Wishart} we discuss sampling of large-$x$ eigenvalue bands.
 
The first moment (mean eigenvalue level) is given by
\begin{align}
  m_1=\frac{1}{N}\,\mathbb E\,[\mathrm{Tr}\,\mathbf C] = \frac{1}{\gamma^2}\sum_{t=1}^T t^{-2\alpha}\,.
\end{align}
As a renormalization condition, we fix the first moment to a constant value $m_1^\star$ which we set to $m_1^\star\equiv1$, without loss of generality.
To implement the renormalization scheme for the Wishart ensemble with power-law variance profile, we fix the mean level across sizes by choosing a running normalization $\gamma^2(T)$ that compensates the growth of the row weights. 
Operationally, one coarse–graining step acts on the data matrix. We remove a fraction of the smallest‑weight rows of $\mathbf Y$ (equivalently, truncate $\mathbf D$ to its top‑left block), retaining $T' := \lfloor T/b \rfloor$ rows with weights $v'_t = t^{-2\alpha}$, $t=1,\dots,T'$.
If only rows are decimated, $N$ is unchanged and the aspect ratio changes as $q' = N/T' = b\,q$.
If a fixed aspect ratio is desired, we decimate columns accordingly, $N' := \lfloor N/b \rfloor$, so that $q' \approx q$. Here, we adopt the latter approach. The normalization $\gamma^2$ is then updated to $\gamma'^2$ so that $m_1'=m_1^\star$, where
\begin{align}
  m_1'=\frac{1}{N'}\mathbb E[\mathrm{Tr}\mathbf C^{(\mathrm{dec})}]\,.
\end{align}
The induced finite–$T$ flow of $\gamma^2$ under the above decimation is shown in Appendix Fig.~\ref{fig:wishart_decimation_flow}, and its large‑$T$ asymptotics is summarized by
\begin{align}
  \gamma^2(T)\propto \sum_{t=1}^T t^{-2\alpha}
  \ \sim\
  \begin{cases}
    \dfrac{T^{1-2\alpha}}{1-2\alpha}, & 0\leq\alpha<1/2\,,\\[6pt]
    \log T, & \alpha=1/2\,,\\[6pt]
    \zeta(2\alpha), & \alpha>1/2 \text{ (as }T\to\infty)\,,
  \end{cases}
  \label{eq:wishart_row_running_gamma}
\end{align}
where $\zeta$ is the Riemann zeta function.
For the RG scale, $s:=\log T$, it is natural to track the running \emph{coupling}
\begin{align}\label{eq:wishart_coupling}
  g(T):=\gamma^{-2}(T)
\end{align}
and define its Beta function by
\begin{align}
  \beta_\mathrm{RG}(g):=\frac{d\log g}{d\log T}\,.
  \label{eq:wishart_row_beta}
\end{align}
The expression is well-defined for all $\alpha$ and quantifies how the strength of the covariance fluctuations changes with scale. For $\alpha<1/2$ one has $\gamma^2(T)\sim T^{1-2\alpha}$, hence $g(T)\sim T^{-(1-2\alpha)}$ and $\beta_\mathrm{RG}\to-(1-2\alpha)<0$. Therefore, the coupling strengthens when the scale is reduced, which is the RG sense of relevance. At $\alpha=1/2$ one has $\gamma^2(T)\sim\log T$, hence $g(T)\sim 1/\log T$ and $\beta_\mathrm{RG}\to0$ with logarithmic corrections, i.e. marginal flow. For $\alpha>1/2$ the normalization saturates, $\gamma^2(T)\to\zeta(2\alpha)$ as $T\to\infty$, so $g(T)\to\zeta(2\alpha)^{-1}$ and $\beta_\mathrm{RG}\to0$, making the heterogeneity irrelevant in the sense that the coupling approaches a size-independent limit. In Appendix~\ref{app:wishart_linearized_flow} we discuss a linearized Beta function that quantifies the approach to the fixed point for $\alpha>1/2$ and recovers a non-zero scaling exponent.

In Figs.~\ref{fig:wishart_simulation_collapse} and~\ref{fig:wishart_de_collapse}, we show a comparison of the raw and collapsed eigenvalue densities, for simulations and solutions of the self-consistent Eqs.~\eqref{eq:wishart_finiteN_G}-\eqref{eq:wishart_finiteN_Sigma}, respectively. Analogous to the Wigner-type ensemble, we visualize the spectra via Cauchy-kernel smoothing, choose a single $\eta_{\mathrm{coll}}$ for all sizes in the collapsed panel, and set $\eta_{\mathrm{raw}}= \gamma^2(T)\,\eta_{\mathrm{coll}}$. We find that eigenvalue densities collapse onto each other, across the different sizes. Differences in the number of density peaks at the right end of the densities are due to better sampling of the power-law profile for larger values of $N$.

With the running normalization $\gamma^2(T)$ given by Eq.~\eqref{eq:wishart_row_running_gamma}, we can define an effective aspect ratio
\begin{align}
  q_{\rm eff}(T):=\frac{N}{\gamma^2(T)}\,.
\end{align}
This ratio plays a role analogous to the Mar\v{c}enko--Pastur parameter $q$ by controlling how well the dominant, large-variance rows are sampled. For $\alpha<1/2$ the growth of $\gamma^2(T)$ yields a non-trivial bulk. At $\alpha=1/2$ the growth is only logarithmic and, in the strict limit, almost all eigenvalues accumulate near zero, while only a vanishing fraction do not approach zero. For $\alpha>1/2$ the normalization saturates and the bulk collapses toward zero.
In this case, the renormalization fixes $m_1$ by a finite number of outliers of size $O(N)$, while the $O(1)$ bulk collapses to zero. We note that at finite $N$, the deterministic fixed-point equation for the resolvent remains quantitatively predictive also for $\alpha>1/2$, provided one compares Cauchy-smoothed densities at fixed smoothing width $\eta$ in the plotted units (and, in simulations, averages over realizations to reduce sampling noise).

Finally, in Appendix~\ref{app:large-N-limit-Wishart}, we provide a derivation of an integral equation for the large-$N$ continuum resolvent and check its correctness through a numerical comparison with the simulated eigenvalue density in Appendix Fig.~\ref{fig:continuum_solution_wishart}.

\section{Discussion}
We used an RG perspective to compare eigenvalue spectra across matrix sizes and implemented it for two structured ensembles—Wigner and Wishart with power‑law variance profiles. These models, or variants thereof, may capture interaction and covariance structure in complex systems, such as heterogeneous synaptic strengths and participation in neural circuits, diffusion and consensus on heterogeneous networks~\cite{OlfatiSaberProcIEEE2007,ChungLuVuPNAS2003,MasudaPorterLambiotte2017}, and sample‑covariance spectra in finance, climate, neuroscience, and genomics where heterogeneous sampling is intrinsic~\cite{Laloux_etal_1999,Plerou1999,vonStorchZwiers1999,CunninghamYu2014,Alter2000}.

The collapsed spectral distribution yields size‑aware dynamical thresholds across different systems. For linear dynamics built from an interaction matrix, stability changes when the right spectral edge crosses the control baseline, and the slowest relaxation time is set by the distance to that edge~\cite{May1972}.
For consensus and diffusion on networks, the Laplacian spectral gap controls mixing times and convergence rates~\cite{OlfatiSaberProcIEEE2007,ChungLuVuPNAS2003}. Finally,
in neural network dynamics, the spectral radius marks the onset of unstable or chaotic activity~\cite{Sompolinsky1988,CunninghamYu2014}. 
Because the RG normalization collapses spectra across sizes, the edge and gap thresholds can be set once at a reference size (in collapsed units) and then applied unchanged across $N$, yielding consistent stability and relaxation criteria. Applying the RG-RMT approach to concrete systems is a natural next step.

To make the recipe operational on data, one should quantify uncertainty—i.e. statistical variability across realizations at fixed $N$ (sampling noise)—and separate it from deterministic finite‑size drift ($1/N$ bias along the RG trajectory), for example by reporting confidence bands for smoothed densities and edge locations (at fixed resolution). Furthermore, sharper finite‑$N$ controls near marginality and at spectral edges would bound these drifts and strengthen guarantees behind collapse and Callan--Symanzik invariance.

We have shown that the RG behaviour is non‑trivial yet analytically tractable. While we do not claim full universality, our evidence provides first hints that after scale fixing, the bulk and edges are largely insensitive to the entry law. A systematic program—varying entry distributions and adding mild structural perturbations—would delineate the empirical domain of attraction of the collapsed spectral distribution and its edge.

Finally, for the two ensembles we have studied, the power‑law structure suggested a natural coarse‑graining procedure via matrix decimation. Extending RG–RMT beyond these cases will likely require model‑specific coarse‑graining rules. Several complementary approaches have been explored on real and synthetic data~\cite{mehta2014exact,koch2018mutual,bradde2017pca,PhysRevLett.123.178103,nicoletti2020scaling,nguyen2025data}, and connecting them to the present RG framework is a promising direction.

\section*{Acknowledgements}
We thank Ilya Nemenman for discussions on RMT in which the idea of using RG for anisotropic matrices via decimating the smallest variance variables emerged and we also thank him for helpful comments on this work.
We would also like to thank Francesca Mastrogiuseppe for discussions.

\appendix

\section{(Anisotropic) local laws}\label{app:local_laws}
In RMT we distinguish spectral resolutions by the scaling of the imaginary part $\eta$ of $z=x+i\eta$: macroscopic resolution keeps $\eta$ fixed as $N\to\infty$; mesoscopic resolution lets $\eta$ shrink with $N$ but still large enough to cover many levels; microscopic resolution probes the mean spacing scale. Concretely:
\begin{align}
  &\eta=O(1)\ \ \text{(macroscopic)}\,,\\
  N^{-1+\varepsilon}\le &\eta \ll 1\ \ \text{(mesoscopic)}\,,\\
  &\eta\lesssim N^{-1}\ \ \text{(microscopic)}\,,
\end{align}
for some fixed $\varepsilon>0$. On mesoscopic scales, \emph{local laws}~\cite{ajanki2017universality,ajanki2019stability} quantitatively control and imply the self‑averaging of resolvents. Recall that $\mathbf G_N^{\mathrm{emp}}$ and $\mathbf G_N$
denote the empirical resolvent and its deterministic equivalent, respectively.
Then averaged statistics concentrate uniformly off the real axis:
\begin{align}
  \frac{1}{N}\,\mathrm{Tr}\,\mathbf G_N^{\mathrm{emp}}(z)\simeq \frac{1}{N}\,\mathrm{Tr}\,\mathbf G_N(z)\,,\\
  \bigg|\frac{1}{N}\,\mathrm{Tr}\big(\mathbf G_N^{\mathrm{emp}}(z)\big)-\frac{1}{N}\,\mathrm{Tr}\big(\,\mathbf G_N(z)\big)\bigg|\ll 1\,,
\end{align}
with typical error scale $(N\eta)^{-1/2}$. In the equations ``$\simeq$'' denotes equality up to controlled, vanishing errors. Or, equivalently, $\,\mathbb E[G_N^{\mathrm{emp}}(z)]=G_N(z)+O((N\eta)^{-1/2})$ on mesoscopic domains. So either route, expectation or deterministic equivalent, yields the same self‑consistent fixed point at the level of accuracy considered here.

The \emph{anisotropic local laws}~\cite{KnowlesYin2017Anisotropic,ajanki2019stability} strengthen this to uniform control of quadratic forms:
\begin{align}
  u^T \mathbf G_N^{\mathrm{emp}}(z)\,v\simeq u^T \mathbf G_N(z)\,v\,,\\
  \sup_{\|u\|=\|v\|=1}\,\big|u^T\big(\mathbf G_N^{\mathrm{emp}}(z)-\mathbf G_N(z)\big)v\big|\ll 1\,,
\end{align}
where $u,v\in\mathbb R^N$ for real ensembles are deterministic unit vectors, $\|u\|=\|v\|=1$, independent of the matrix randomness. The anisotropic local law provides uniform control of the quadratic form $u^T\mathbf G_N(z)v$ over all such $u$ and $v$.
The law holds on mesoscopic domains
with possible model‑dependent modifications near spectral edges or singularities. Physically, $\eta$ sets a mesoscopic spectral coarse‑graining scale: a window of width $\eta$ averages over $\sim N\,\eta\,\rho(x)$ levels (eigenvalues), making resolvent observables effectively deterministic on that scale.
In this sense the imaginary part $\eta$ acts as a spectral coarse‑graining parameter: increasing $\eta$ broadens the Cauchy kernel and averages finer‑scale fluctuations, while decreasing $\eta$ moves one toward microscopic resolution. As $N$ grows, self‑averaging random observables converge to their deterministic equivalents, with vanishing fluctuations of order $(N\eta)^{-1/2}$.

\section{Derivation of the self-consistent equation for Wigner-type ensemble}\label{app:Wigner-type-eqn}
The goal is to derive a site-level Dyson form for the diagonal resolvent. We start from the resolvent identity
\begin{align}
  (z\mathbf I_N-\mathbf X_N)\,\mathbf G^\mathrm{emp}_N(z)=\mathbf I_N\,,\quad z=x+i\eta\,,\ \eta>0\,.
\end{align}
The diagonal entries satisfy
\begin{align}
  z\,G^\mathrm{emp}_{ii} - X_{ii}\,G^\mathrm{emp}_{ii} - \sum_{k\ne i} X_{ik}\,G^\mathrm{emp}_{ki} = 1\,,
  \label{eq:res_identity_site_final}
\end{align}
where for ease of notation we have dropped the label $N$. In the following subsection we also omit the label ``emp" from the empirical resolvent to avoid clutter of notation.  

\subsection{Integration by parts}
We derive expressions for the expectation values of the different terms in Eq.~\eqref{eq:res_identity_site_final}. We assume Gaussian entries and use integration by parts (IBP), which gives the standard relation,
$\mathbb E[X_{ab}\,F(\mathbf X)] = \mathrm{Var}(X_{ab})\,\mathbb E\big[\partial F/\partial X_{ab}\big]$.
The derivative of the resolvent is
\begin{align}
  \frac{\partial \mathbf G}{\partial X_{ab}}
  = \mathbf G\,\frac{\partial \mathbf X}{\partial X_{ab}}\,\mathbf G\,,
\end{align}
and for symmetric $\mathbf X$:
\begin{align}
    \frac{\partial \mathbf X}{\partial X_{ik}}= \mathbf E_{ik}+\mathbf E_{ki}\,, \quad \frac{\partial \mathbf X}{\partial X_{ii}}=\mathbf E_{ii}\,,
\end{align}
with $i\neq k$ and $(\mathbf E_{ab})_{uv}=\delta_{au}\delta_{bv}$. In particular
\begin{align}
  \frac{\partial G_{ii}}{\partial X_{ii}}
  = (\mathbf G\,\mathbf E_{ii}\,\mathbf G)_{ii}
  = G_{ii}^2\,.
\end{align}

Let $s_{ik}$ denote the entrywise variance profile defined in Eq.~\eqref{eq:var_matrix}.
Taking the expectation value of Eq.~\eqref{eq:res_identity_site_final} and applying Gaussian IBP yields for the third term on the left-hand side:
\begin{align}\label{eq:XikGki}
    \mathbb E\big[X_{ik}\,G_{ki}\big]
    = s_{ik}\,\mathbb E\Big[\frac{\partial G_{ki}}{\partial X_{ik}}\Big]
    = s_{ik}\,\mathbb E\big[ G_{kk}G_{ii}+G_{ki}^2\big]\,,
\end{align}
with $i\neq k$. Likewise, for the second term:
\begin{align}\label{eq:XiiGii}
    \mathbb E\big[X_{ii}\,G_{ii}\big]
    = s_{ii}\,\mathbb E\Big[\frac{\partial G_{ii}}{\partial X_{ii}}\Big]
    = s_{ii}\,\mathbb E\big[ G_{ii}^2\big]\,.
\end{align}

\subsection{Mean-field closure}

We now pass from the empirical identity to a deterministic site equation using self‑averaging, cf. Appendix~\ref{app:local_laws}. Taking expectations in Eq.~\eqref{eq:res_identity_site_final} gives
\begin{align}\label{eq:identy_expectation}
  z\,\mathbb E[G^\mathrm{emp}_{ii}(z)] - \sum_{k=1}^N \mathbb E\big[X_{ik}\,G^\mathrm{emp}_{ki}(z)\big] = 1\,.
\end{align}

On mesoscopic scales ($\eta\ge N^{-1+\varepsilon}$) off the real axis, the anisotropic local law implies self‑averaging of diagonal resolvent entries and suppression of off‑diagonals, uniformly in $i$ and $k$:
\begin{align}
  G^{\mathrm{emp}}_{ii}(z)&=G_{ii}(z)+O\big((N\eta)^{-1/2}\big)\,,\\
  G^{\mathrm{emp}}_{ki}(z)&=O\big((N\eta)^{-1/2}\big)\,,
\end{align}
with $i\neq k$.
Hence in Eqs.~\eqref{eq:XikGki}-\eqref{eq:XiiGii} mixed products factorize up to local‑law errors and off‑diagonal squares are negligible:
\begin{align}
  \mathbb E\big[G^{\mathrm{emp}}_{kk}G^{\mathrm{emp}}_{ii}\big]
  &= G_{kk}G_{ii}+O\big((N\eta)^{-1/2}\big)\,,\\
  \mathbb E\big[(G^{\mathrm{emp}}_{ki})^2\big]&=O\big((N\eta)^{-1}\big)\,.
\end{align}
Therefore, for Eqs.~\eqref{eq:XikGki}-\eqref{eq:XiiGii} we have the approximations:
\begin{align}
  \mathbb E\big[X_{ik}\,G^{\mathrm{emp}}_{ki}\big]\simeq s_{ik}\,G_{kk}\,G_{ii}\,,\quad
  \mathbb E\big[X_{ii}\,G^{\mathrm{emp}}_{ii}\big]\simeq s_{ii}\,G_{ii}^2\,.
\end{align} 
Substitution into the site equation~\eqref{eq:identy_expectation}, and dividing by $G_{ii}$, yields the closed equation for the deterministic equivalent
\begin{align}
  \frac{1}{G_{ii}(z)} = z - \sum_{k=1}^N s_{ik}\,G_{kk}(z)\,.
\end{align}
Thus the self‑energy closes as
\begin{align}
  \Sigma_{ii}(z) := \sum_{k=1}^N s_{ik}\,G_{kk}(z)\,,\quad \boldsymbol\Sigma(z)=\mathbf S\,\mathbf G_N(z)\,,
\end{align}
and, with $s_{ik}$ from Eq.~\eqref{eq:var_matrix} we obtain the form
\begin{align}
  \frac{1}{G_i}= z - \frac{\sigma^2}{\gamma}\Big(w_i \sum_{k=1}^N w_k G_k + w_i^2 G_i\Big)\,,
\end{align}
where we have defined $G_i:=G_{ii}$.
Using the definition of the coupling and variance-weighted resolvent in Eq.~\eqref{eq:order_param},
this becomes the on‑site Dyson form
\begin{align}\label{eq:site_dyson_phys}
  \frac{1}{G_i}= z - c\big(w_i\,u(z)+w_i^2 G_i\big)\,,
\end{align}
with the physical branch fixed by $\Im\,G_i(z)<0$.

\section{Fixed-point algorithm for Wigner-type ensemble}\label{app:algorithm-wigner}
We solve for the variance-weighted mean field $u(z)$ by fixed-point iteration (indexed by superscript $n$) at $z=x+i\,\eta$ with small $\eta>0$~\cite{ortega2000iterative}:

\begin{algorithm}[H]
\caption{Fixed-point iteration for $u(z)$ and $G_N(z)$ (Wigner-type ensemble)}\label{algo:algorithm_1}
\KwIn{$z=x+i\,\eta$ with $\eta>0$, tolerance $\varepsilon>0$, damping $\theta\in(0,1]$}
\KwInit{$w_j\leftarrow j^{-\alpha}$, $c\leftarrow \sigma^2/\gamma$, $W_j\leftarrow c\,w_j^2$; $u^{(0)} \leftarrow -i\,c_0\sum_{j=1}^N w_j$ with $c_0\in(0,1)$; $n \leftarrow 0$}
\text{(Warm start:)}\\
\For{$j \gets 1$ \KwTo $N$}{
  Compute $G_j^{(0)}$ from the quadratic $W_j (G_j^{(0)})^2 - (z - c\,w_j\,u^{(0)}) G_j^{(0)} + 1 = 0$, choosing the branch with $\Im\,G_j^{(0)}<0$;
}
$u^{(1)} \leftarrow \sum_j w_j\,G_j^{(0)}$;\quad $n \leftarrow 1$;\\
\Repeat{$\lvert u^{(n)}-u^{(n-1)}\rvert < \varepsilon$}{
  \For{$j \gets 1$ \KwTo $N$}{
    Compute $G_j^{(n)}$ from $W_j (G_j^{(n)})^2 - (z - c\,w_j\,u^{(n)}) G_j^{(n)} + 1 = 0$, choose $\Im\,G_j^{(n)}<0$;
  }
  $u^{(n+1)} \leftarrow (1-\theta)\,u^{(n)} + \theta \sum_j w_j\,G_j^{(n)}$;\\
  $n \leftarrow n+1$;
}
$G_N(z) \leftarrow \frac{1}{N}\sum_j G_j^{(n)}$;\\
$\rho_{N,\eta}(x) \leftarrow -\,\frac{1}{\pi}\,\Im\,G_N(x+i\,\eta)$;
\end{algorithm}

\section{Visualizing the eigenvalue densities}\label{app:visualization}
We visualize spectra via Cauchy–kernel smoothing~\cite{potters2020first}. For the empirical resolvent
\begin{align}
  \rho^{\mathrm{emp}}_{N,\eta}(x)
  &:= -\,\frac{1}{\pi}\,\Im\,G^{\mathrm{emp}}_N(x+i\eta)\nonumber\\
  &= \frac{1}{\pi}\Big[\frac1N\sum_{j=1}^N \frac{\eta}{(x-\lambda_j)^2+\eta^2}\Big]\,,
\end{align}
where $\{\lambda_j\}$ are the empirical eigenvalues. For the deterministic equivalent $G_N$ we define the smoothed density
\begin{align}
  \rho_{N,\eta}(x) := -\,\frac{1}{\pi}\,\Im\,G_N(x+i\eta)\,,
\end{align}
which is the same Cauchy smoothing but without an eigenvalue sum. In both cases $\eta$ is the half–width (resolution) of the Cauchy kernel and fixes the resolution scale: features narrower than $\eta$ are averaged out, while broader structures are preserved. 
For an $N\times N$ matrix, the microscopic mean level spacing is $\Delta(x)\sim 1/(N\,\rho_{N,\eta}(x))$.
Choosing $\eta$ with $\Delta(x)\ll \eta \ll O(1)$ defines a mesoscopic resolution: the Cauchy kernel averages over $\sim N\,\eta\,\rho_{N,\eta}(x)$ eigenvalues, suppressing level-to-level noise while preserving $O(1)$ variations of the density. On this scale, the empirical resolvent concentrates around its deterministic equivalent and the Cauchy–smoothed empirical density tracks deterministic eigenvalue density.

In collapsed curves, the eigenvalues are rescaled by the running normalization
$\gamma(N)$ that fixes the second moment, according to $x=\sqrt{\gamma(N)}\,y$ (the
spectral bandwidth scales like $\sqrt{\gamma(N)}$ in raw units). To compare
raw and collapsed curves with the same spectral resolution, the
smoothing width must transform covariantly under this change of units:
since $dx=\sqrt{\gamma(N)}\,dy$, keeping the kernel width fixed in $y$–units
implies
\begin{align}
  \eta_{\mathrm{raw}} = \sqrt{\gamma(N)}\,\eta_{\mathrm{coll}}\,.
\end{align}
Thus, we choose the same $\eta_{\mathrm{coll}}$ for all values $N$, and set $\eta_{\mathrm{raw}}$ via the above relation. Using a constant value of $\eta$ for all raw curves would lead to under- or over‑smoothing at large or small
$N$, and break comparability across
sizes.

On mesoscopic scales the smoothed density $\rho_{N,\eta}$ is strictly positive outside the true support due to the Cauchy kernel. This “tail” is a visualization artifact of the width set by $\eta$ and decays as $\sim \eta/(x-\text{edge})^2$. We therefore use a single mesoscopic $\eta$ across sizes and, for fair comparison, apply the same kernel to empirical spectra and to the MDE solution.

\section{Large-$N$ behaviour of the generalized harmonic sum}\label{app:large_N}
We define the generalized harmonic sum
\begin{align}
  H_N(\theta) := \sum_{j=1}^N j^{-\theta}\,.\label{eq:HN_def}
\end{align}
The large-$N$ behaviour of the generalized harmonic sum is classical (see, e.g., Ch.~3 of~\cite{ApostolANT}) and has three regimes:
\begin{align}
  H_N(\theta) &= \frac{N^{1-\theta}}{1-\theta}+ O(1)\,, && 0\le \theta<1\,,\label{eq:HN_lt1}\\
  H_N(1) &= \log N+ \gamma_\mathrm{E}+ O(N^{-1})\,, && \theta=1\,,\label{eq:HN_eq1}\\
  H_N(\theta) &= \zeta(\theta)+ O\!\big(N^{-(\theta-1)}\big)\,, && \theta>1\,,\label{eq:HN_gt1}
\end{align}
where $\gamma_\mathrm{E}= 0.57721...$ is the Euler--Mascheroni constant and $\zeta$ is the Riemann zeta function
\begin{align}
    \zeta(\theta):=\sum_{j=1}^\infty j^{-\theta}\,,\quad \theta>1\,.
\end{align}
Thus, the dominant scaling of the generalized harmonic number is
\begin{align}
  H_N(\theta) \sim\begin{cases}
    N^{1-\theta}\,, & \theta<1\,,\\
    \log N\,, & \theta=1\,,\\
    1\,, & \theta>1\,.
  \end{cases}\label{eq:HN_scaling}
\end{align}

\begin{figure}[t]
    \centering
    \includegraphics[width=0.95\linewidth]{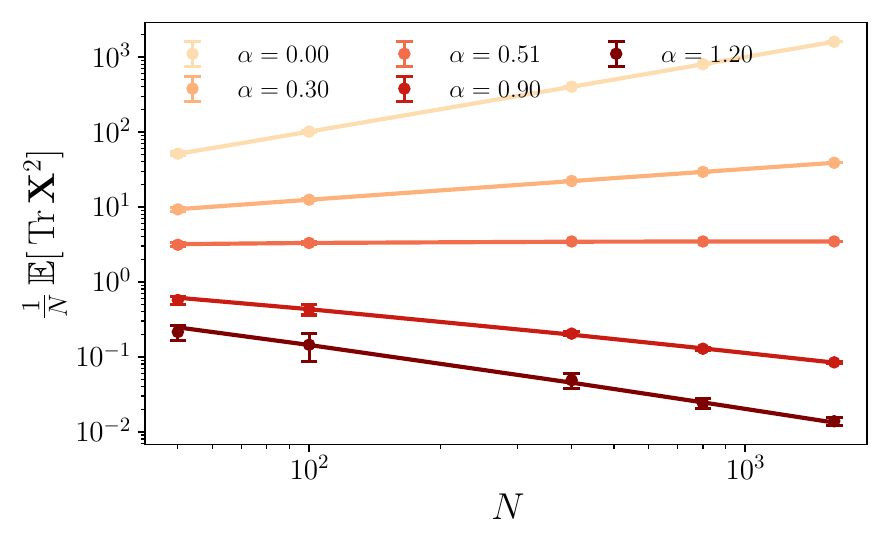}
    \caption{{\bf Finite-size scaling of the second moment for representative exponents $\alpha$.} Solid lines: exact formula in Eq.~\eqref{eq:second_moment}. Markers: Simulation estimates from eight independent trials. Dashed guides: predicted slopes.}
    \label{fig:m2_scaling}
\end{figure}

\begin{figure*}[t]
\centering
\includegraphics[width=16.8cm]{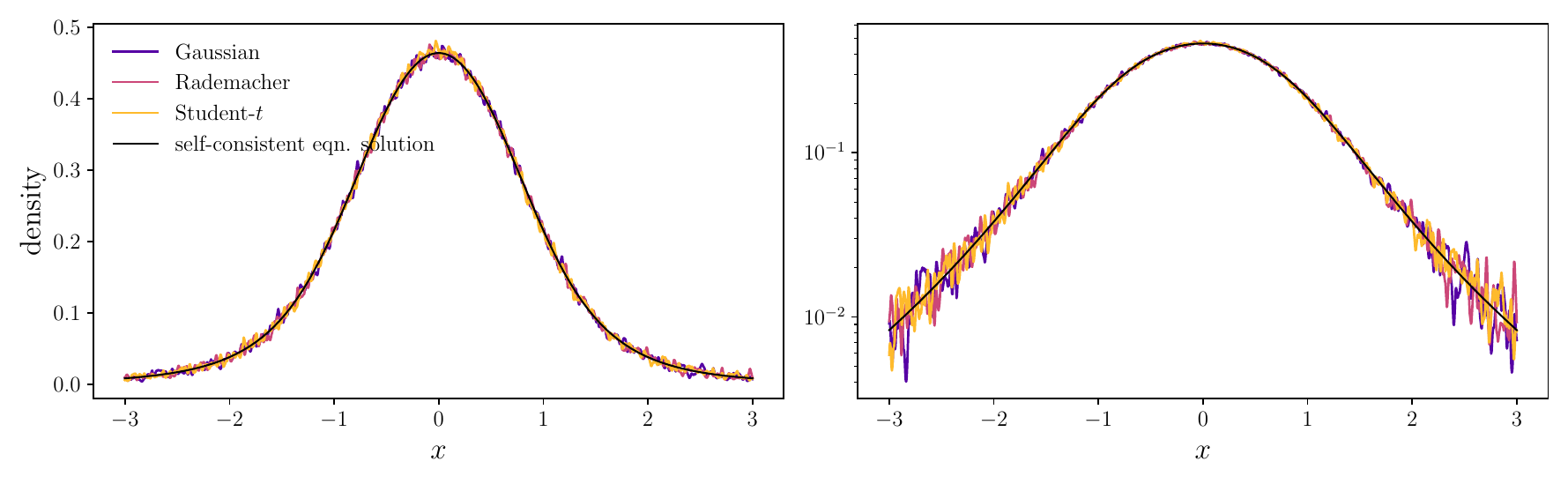}
\caption{{\bf Entry-law universality.} Eigenvalue densities from simulations with Gaussian (purple), Rademacher (magenta), and Student-$t$ with $\nu=4$ (yellow)  matrix entries collapse onto the solution of the self-consistent equation (black) under the RG normalization $\gamma(N)$. Linear-linear scale (left) and log-linear scale (right). Parameters are $N=600$, $\alpha=0.60$, and $\eta_\mathrm{coll}=10^{-2}$ and ensembles are simulated with $20$ independent trials.}
\label{fig:universality_demos}
\end{figure*}

\section{Large-$N$ limit and the continuum fixed point}\label{app:large-N-limit-Wigner}
We start from the finite-$N$ self-consistent Eq.~\eqref{eq:quad_Gj} and pass to a large-$N$ continuum limit~\cite{AGZ2010}. This is useful because it transforms the discrete fixed-point equation into a closed integral equation with a rank‑one self‑energy, providing a universal description and analytic control of the spectrum. As before, we choose the running coupling $c(N)=\sigma^2/\gamma(N)$ such that the second moment in Eq.~\eqref{eq:second_moment_X} is scale‑fixed by the RG condition in Eq.~\eqref{eq:rg_condition}. 

We reparameterize indices and weights and assume a smooth resolvent profile:
\begin{align}\label{eq:cont_relations}
  y_j := \frac{j}{N}\,, \quad
  w_j = j^{-\alpha} = N^{-\alpha} y_j^{-\alpha}\,, \quad
  G_j(z) \simeq G(y_j;z)\,.
\end{align}
We note that $y_j=j/N\in(0,1]$ such that the coordinate $y_j$ becomes continuous in the large-$N$ limit.
The profile $G(y;z)$ is the leading‑order dependence of the diagonal resolvent on the continuous position $y$ induced by the slowly varying weights $w_j$ in mean‑field self‑energy form.

First, we treat the term capturing the interaction with the variance-weighted mean field in Eq.~\eqref{eq:quad_Gj} which, using the above relations in Eq.~\eqref{eq:cont_relations}, we write as
\begin{align}\label{eq:continuum_cwu}
    cw_j\sum_{k=1}^N w_k G_k &= c(N)N^{1-2\alpha}\big[\frac1N\sum_{k=1}^N y_j^{-\alpha}\, y_k^{-\alpha}\, G_k\big]\,.
\end{align}
We define the continuum amplitude
\begin{align}
  h_\alpha := \lim_{N\to\infty} \big[c(N)\,N^{1-2\alpha}\big]
  = \begin{cases}
      (1-\alpha)^2\,, & 0<\alpha<1\,,\\[4pt]
      0\,,            & \alpha\geq1\,.
    \end{cases}
\end{align}
At $\alpha=1$, the amplitude $h_\alpha$ vanishes due to log-suppression.
Hence the Riemann sum in Eq.~\eqref{eq:continuum_cwu} becomes an integral and, overall,
\begin{align}
  c\,w_j u \;\xrightarrow[N\to\infty]{}\;
  h_\alpha \int_0^1 y^{-\alpha}y'^{-\alpha} G(y';z)\,dy'\,,
\end{align}
and the rank‑one field vanishes in the continuum limit for $\alpha\geq1$.

Next, we treat the local term in Eq.~\eqref{eq:quad_Gj} in an analogous manner and find that it vanishes in the large-$N$ limit. Specifically:
\begin{align}
  c\,w_j^2 G_j 
  &= c(N)N^{1-2\alpha}\frac{y_j^{-2\alpha}}{N}G_j
\end{align}
and in the large-$N$ limit
\begin{align}\label{eq:aj_scaling_cont}
  c\,w_j^2 G_j \xrightarrow[N \rightarrow \infty]{} \frac{h_\alpha}{N}\,y^{-2\alpha}\,G(y;z)\,.
\end{align}
Thus, $c\,w_j^2 G_j=O(N^{-1})$ uniformly in $x$ for $z=x+i\eta$ with fixed $\eta>0$ (log‑corrected at $\alpha=1$)\,.
Since $G(y;z)=O(1)$ off the real axis, the local term contribution vanishes in the large‑$N$ limit.

\begin{figure}[t]
\centering
\includegraphics[width=8.4cm]{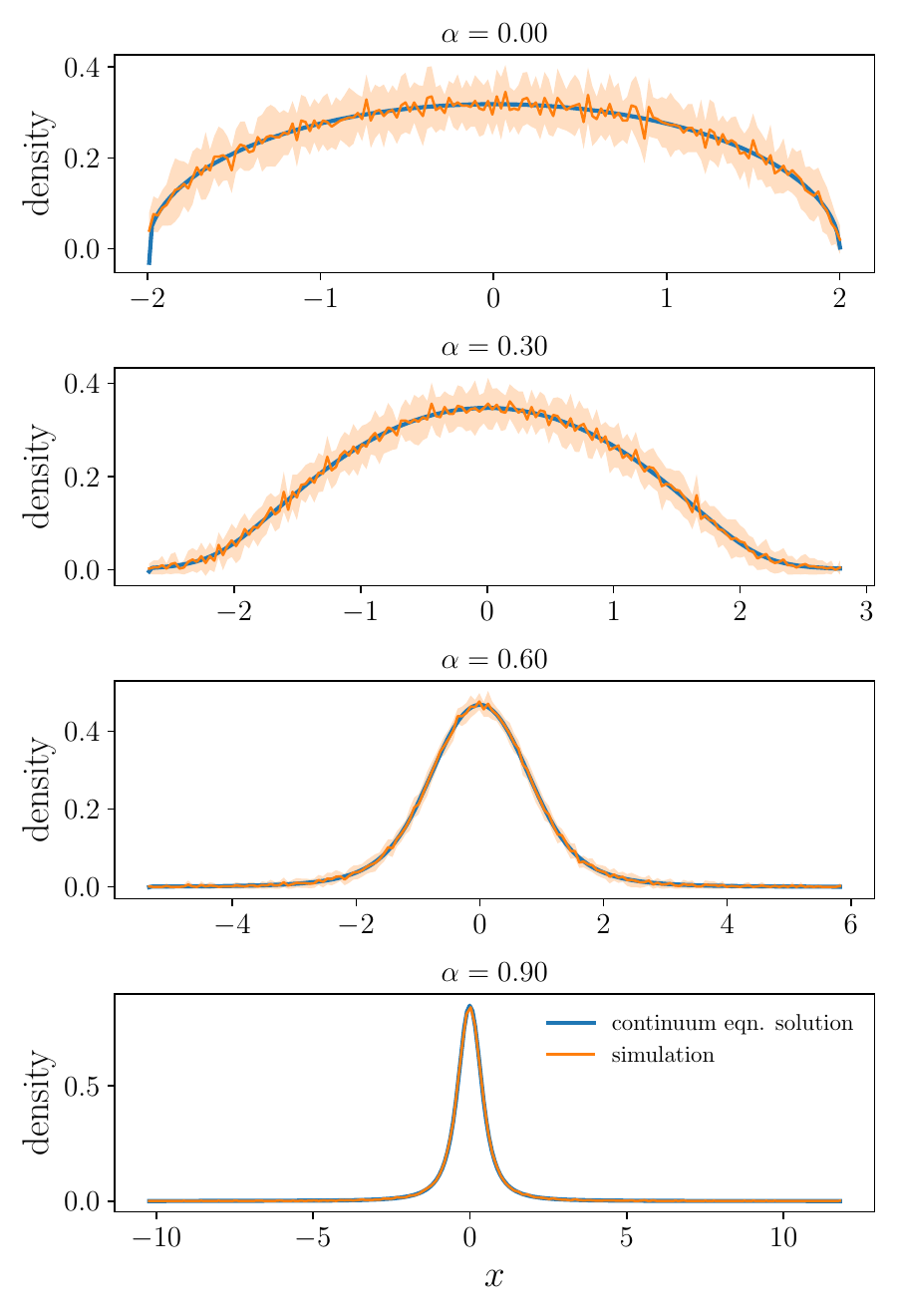}
\caption{{\bf Comparison of continuum equation solution and simulation.} Plots show the continuum limit solution (blue) and the simulation with mean and one standard deviation band (orange) for $\alpha=0$, $0.3$, $0.6$, and $0.9$. Simulation curves are obtained with $N=600$ and $20$ independent trials.}\label{fig:continuum_solution}
\end{figure}

Thus, as the continuum limit form of Eq.~\eqref{eq:quad_Gj}, we find the Dyson integral equation 
\begin{align}\label{eq:dyson_kernel}
    G(y;z) &= \frac{1}{z - \Sigma(y;z)}
\end{align}
with self-energy
\begin{align}\label{eq:selfenergy_kernel}
    \Sigma(y;z) &:= \int_0^1 K(y,y')\,G(y';z)\,dy'
\end{align}
and the kernel:
\begin{align}\label{eq:kernel_rank1}
  K(y,y') &:= h_\alpha\,y^{-\alpha}y'^{-\alpha}\,.
\end{align}
This form of the Dyson equation is consistent with the abstract Dyson equation point of view of~\cite{alt2020dyson}.
Since the kernel factorizes, we can define
\begin{align}
    \Sigma(y;z)&=y^{-\alpha}S(z)\,,\quad S(z):=h_\alpha\,U(z)\,,\nonumber\\
U(z)&:=\int_0^1 y^{-\alpha}G(y;z)\,dy\,.
\end{align}
Then
\begin{align}
    G(y;z)=\frac{1}{z-\Sigma(y;z)}=\frac{1}{z\big(1-\phi(z)\,y^{-\alpha}\big)}\,,
\end{align}
where in the last equality we have defined the normalized self‑energy amplitude
\begin{align}
    \phi(z):=\frac{S(z)}{z}=\frac{h_\alpha}{z}\,U(z)\,.
\end{align}
Using the definition of $U(z)$, we obtain the scalar fixed point
\begin{align}
    \phi(z)-\frac{h_\alpha}{z^2}\,Q_\alpha\!\big(\phi(z)\big)=0\,,\quad
    Q_\alpha(\phi):=\int_0^1\frac{dy}{y^{\alpha}-\phi}\,.
\end{align}
The mean resolvent follows as
\begin{align}
    \overline{G}(z)=\int_0^1 G(y;z)\,dy
    = \frac{1}{z}\,\big(1+\phi\,Q_\alpha(\phi)\big)\,.
\end{align}
At the marginal value $\alpha=1/2$ the RG fixed point is non‑trivial: the continuum kernel has $h_{1/2}=1/4$, so the Dyson self‑energy does not vanish and the resolvent is not the free resolvent, $1/z$. On the other hand, for $\alpha>1$
\begin{align}
    \Sigma= 0\,,\quad \text{and}\quad \overline G(z)=\frac{1}{z}\,,
\end{align}
i.e. the continuum self-energy vanishes and the bulk of the eigenvalues collapses toward zero, while a finite number of $O(1)$--outliers (associated with small indices) remain. These outliers are not self‑averaging.
In Fig.~\ref{fig:continuum_solution}, we show a comparison of the continuum limit form and the eigenvalue density obtained from simulated ensembles and find a near-perfect match.

Within the class of centered, independent entries with finite variance and standard moment bounds, the large-$N$ deterministic limit supplied by the matrix Dyson equation depends on the model only through its variance operator; see, e.g.,~\cite{ajanki2017universality,ajanki2019stability,KnowlesYin2017Anisotropic}. Under our RG normalization that fixes $m_2$ and for the power‑law weights, the variance kernel factorizes, so the continuum self‑energy is rank‑one and its amplitude is $h_\alpha$. Consequently, the bulk limiting density depends on $\alpha$ only through $h_\alpha$, and is insensitive to higher cumulants at leading order on mesoscopic scales. 

We note that the above statement is restricted to mean‑field regimes with finite variance. Subleading corrections, edge behaviour, and finite‑$N$ fluctuations can be model‑dependent, and heavy‑tailed (infinite‑variance), sparse, or banded ensembles also lie outside this scope. More generally, (non‑factorizable) variance profiles would lead to higher‑rank kernels and different fixed points.

\section{Derivation of the self-consistent equation for Wishart-type ensemble}\label{app:wishart_finiteN_cavity}
\begin{figure}[t]
    \centering
    \includegraphics[width=0.95\linewidth]{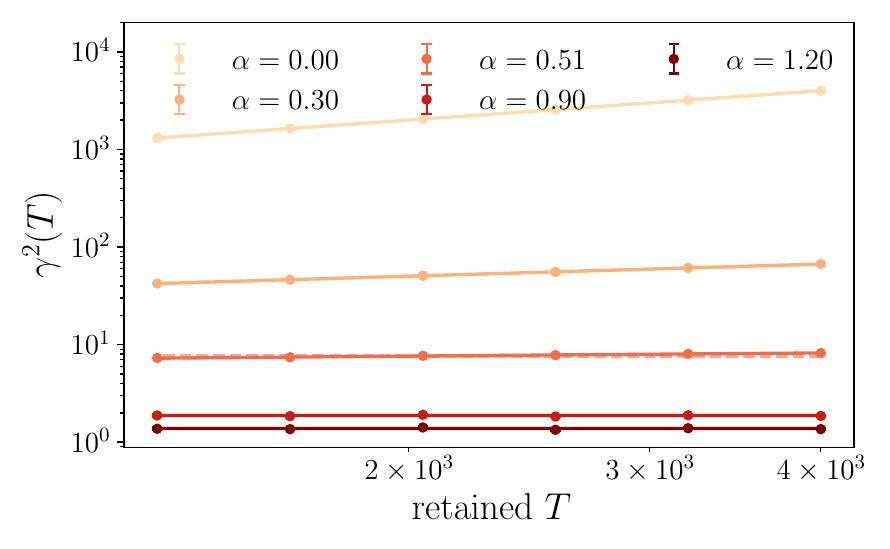}
    \caption{{\bf Normalization flow with decimation.} Decimation flow of $\gamma^2(T)$ with retained size $T$ after removing a fixed fraction of largest indices at each step, while the first moment is kept at unity. Solid lines: exact prediction. Markers with error bars show the simulation mean and standard deviation computed from six independent trials.}
    \label{fig:wishart_decimation_flow}
\end{figure}
We consider $\mathbf Z\in\mathbb R^{T\times N}$ with i.i.d. centered unit-variance entries, $\mathbf D=\mathrm{diag}(v_1,\dots,v_T)$ with $v_t=t^{-2\alpha}$, and define
\begin{align}
    \mathbf Y=\mathbf D^{1/2}\mathbf Z\,,\quad
    \mathbf C=\frac{1}{\gamma^2}\,\mathbf Y^T \mathbf Y=\frac{1}{\gamma^2}\,\mathbf Z^T \mathbf D\,\mathbf Z\,.
\end{align}

\subsection{Row decomposition and rank–one update}
We write the $t$-th row of $\mathbf Z$ as $z_t\in\mathbb R^N$ and set
\begin{align}
  b_t := \frac{\sqrt{v_t}}{\gamma}\,z_t \sim \mathcal N\!\big(0,\,(v_t/\gamma^2)\,\mathbf I_N\big)\,,
\end{align}
so that the covariance matrix is a sum of rank–one outer products:
\begin{align}\label{eq:C-bt}
    \mathbf C=\sum_{t=1}^T b_t b_t^T\,.
\end{align}
Starting from the resolvent identity
\begin{align}
(z\,\mathbf I_N-\mathbf C)\,\mathbf G_N(z) &= \mathbf I_N\,,
\end{align}
and taking the normalized trace, we obtain
\begin{align}
    z\,G_N(z) - \frac{1}{N}\sum_{t=1}^T b_t^{\!T}\,\mathbf G_N(z)\,b_t=1\,,
\end{align}
with $G_N(z):=\tfrac{1}{N}\mathrm{Tr}\,\mathbf G_N(z)$.
We remove the $t$'th row from $\mathbf Z$ and define the \emph{leave-one-out} quantities
\begin{align}
  \mathbf C^{(t)} := \sum_{s\ne t} b_s b_s^T\,,\quad
  \mathbf G_N^{(t)}(z) := (z\,\mathbf I_N - \mathbf C^{(t)})^{-1}\,.
\end{align}
By the Sherman-Morrison formula~\cite{10.1214/aoms/1177729698,potters2020first}:
\begin{align}
    (\mathbf A + u v^T)^{-1}
    = \mathbf A^{-1} - \frac{\mathbf A^{-1} u v^T \mathbf A^{-1}}{1 + v^T \mathbf A^{-1} u}
\end{align}
with $\mathbf A^{-1}=\mathbf G_N^{(t)}$ and $uv^T=\mathbf G_N^{-1}-{\mathbf G_N^{(t)}}^{-1}=-b_tb_t^T$, we obtain the rank‑one update of the resolvent and its trace:
\begin{align}
    \mathbf G_N(z)
    &= \mathbf G_N^{(t)}(z)+\frac{\mathbf G_N^{(t)}(z)\,b_t b_t^{\!T}\,\mathbf G_N^{(t)}(z)}{1 - b_t^{\!T}\mathbf G_N^{(t)}(z)\,b_t}
    \\[4pt]
    \Leftrightarrow\quad
    b_t^{\!T}\mathbf G_N(z)\,b_t
    &= b_t^{\!T}\mathbf G_N^{(t)}(z)\,b_t+\frac{\big[b_t^{\!T}\mathbf G_N^{(t)}(z)\,b_t\big]^2}{1 - b_t^{\!T}\mathbf G_N^{(t)}(z)\,b_t}\,,
\end{align}
where we have used Eq.~\eqref{eq:C-bt}. Hence, we obtain
\begin{align}\label{eq:bTGb}
    b_t^{\!T}\mathbf G_N(z)\,b_t &= \frac{b_t^{\!T}\mathbf G_N^{(t)}(z)\,b_t}{1 - b_t^{\!T}\mathbf G_N^{(t)}(z)\,b_t}\,.
\end{align}

\subsection{Conditional Gaussian averaging}
The random objects we average are the quadratic forms in the numerator and denominator on the right-hand side of Eq.~\eqref{eq:bTGb}.
We condition on the \emph{leave–one–out} (minor) resolvent $\mathbf G_N^{(t)}$ (i.e., treat $\mathbf C^{(t)}$ as fixed), and note that $b_t$ is independent of $\mathbf C^{(t)}$. Since $b_t\sim \mathcal N(0,(v_t/\gamma^2)\mathbf I)$ and $\mathbf G_N^{(t)}$ is fixed under the conditional expectation, the standard Gaussian contraction gives
\begin{align}
  \mathbb E_{b_t}\!\big[b_t^T \mathbf G_N^{(t)} b_t \,\big|\, \mathbf G_N^{(t)}\big]
  = \mathrm{Tr}\!\big(\mathrm{Cov}(b_t)\,\mathbf G_N^{(t)}\big)
  = \frac{v_t}{\gamma^2}\,\mathrm{Tr}\,\mathbf G_N^{(t)}(z)\,.
\end{align}
Thus, under the conditional expectation, the quadratic form reduces to a trace against the leave–one–out resolvent, and we have
\begin{align}
    \mathbb E_{b_t}\!\big[b_t^{\!T}\mathbf G_N(z)\,b_t \,\big|\, \mathbf G_N^{(t)}\big]
    = \frac{\frac{v_t}{\gamma^2}\,\mathrm{Tr}\,\mathbf G_N^{(t)}(z)}{1 - \frac{v_t}{\gamma^2}\,\mathrm{Tr}\,\mathbf G_N^{(t)}(z)}\,.
\end{align}

\subsection{Mean–field replacement and equation closure}
The anisotropic local law (cf. Appendix~\ref{app:local_laws} and~\cite{KnowlesYin2017Anisotropic,ajanki2019stability}), asserts that, uniformly for $z=x+i\eta$ with $\eta>0$, leave–one–out resolvents $\mathbf G^{(t)}$ and full resolvents $\mathbf G_N$ differ by subleading corrections in quadratic forms and traces:
\begin{align}
  \frac{1}{N}\mathrm{Tr}\,\mathbf G_N^{(t)}(z)
= G_N(z) + O\big((N\eta)^{-1/2}\big)\,.
\end{align}
Substituting into the normalized trace identity yields, up to vanishing errors
\begin{align}
    z\,G_N(z) - \frac{1}{N}\sum_{t=1}^T \frac{\big(\tfrac{v_t}{\gamma^2}\big)\,N\,G_N(z)}{1 - \big(\tfrac{v_t}{\gamma^2}\big)\,N\,G_N(z)}=1\,,
\end{align}
or, equivalently, the closed fixed point
\begin{align}
    G_N(z) &=\frac{1}{z - \Sigma_N(z)}\,,\\[4pt]
    \Sigma_N(z) &= \frac{1}{\gamma^2}\sum_{t=1}^T \frac{v_t}{1 - \big(N/\gamma^2\big)\,v_t\,G_N(z)}\,.
\end{align}
These are the finite‑$N$ fixed-point equations for the deterministic resolvent. Empirical resolvents concentrate around it with local‑law accuracy. The correction term is of order $(N\eta)^{-1/2}$. The same fixed-point equations can also be derived from the Silverstein–Dozier framework for sample covariances~\cite{BaiSilverstein_book,SilversteinBai1995,DozierSilverstein2007}.

\subsection{Fixed-point algorithm for the Wishart-type ensemble}\label{app:algo_2}
We solve the self-consistent equations for the Wishart-type ensemble Eqs.~\eqref{eq:wishart_finiteN_G}--\eqref{eq:wishart_finiteN_Sigma}, with the following fixed-point algorithm~\cite{ortega2000iterative}:
\begin{algorithm}[H]
\caption{Fixed-point iteration for $G_N(z)$ (Wishart-type ensemble)}\label{algo:algorithm_2}
\KwIn{$z=x+i\,\eta$ with $\eta>0$; $\alpha, N, T, \gamma^2$; tolerance $\varepsilon>0$; damping $\theta\in(0,1]$; stabilization $\varepsilon_{\rm stab}>0$}
\KwInit{$v_t \leftarrow t^{-2\alpha}$ for $t=1,\dots,T$; $G^{(0)} \leftarrow 1/z$; $n \leftarrow 0$}
\text{(Warm start:)}
Compute $\displaystyle \Sigma^{(0)} \leftarrow \frac{1}{\gamma^2}\sum_{t=1}^T \frac{v_t}{1 - (N/\gamma^2)\,v_t\,G^{(0)} + i\,\varepsilon_{\rm stab}}$; set $\displaystyle G^{(1)} \leftarrow \frac{1}{z - \Sigma^{(0)}}$; set $n \leftarrow 1$;\\
\Repeat{$\big|G^{(n)}-G^{(n-1)}\big|<\varepsilon$}{
  Compute $\displaystyle \Sigma^{(n)} \leftarrow \frac{1}{\gamma^2}\sum_{t=1}^T \frac{v_t}{1 - (N/\gamma^2)\,v_t\,G^{(n)} + i\,\varepsilon_{\rm stab}}$;\\
  Set $\displaystyle \widetilde G^{(n+1)} \leftarrow \frac{1}{z - \Sigma^{(n)}}$;\\
  $G^{(n+1)} \leftarrow (1-\theta)\,G^{(n)} + \theta\,\widetilde G^{(n+1)}$;\\
  $n \leftarrow n+1$;
}
$G_N(z)\leftarrow G^{(n)}$;\quad $\rho_{N,\eta}(x)\leftarrow -\,\dfrac{1}{\pi}\Im\,G_N(z)$;
\end{algorithm}

\section{Linearized RG flow in the saturating regime ($\alpha>1/2$)}\label{app:wishart_linearized_flow}

In the main text we define the RG Beta function in terms of the running coupling $g(T):=\gamma^{-2}(T)$, cf. Eq.~\eqref{eq:wishart_coupling}. For $\alpha>1/2$ the renormalization condition implies that $\gamma^2(T)$ saturates and hence $g(T)$ approaches a nonzero fixed point $g_\star$. In this situation the Beta function $\beta_\mathrm{RG}(g)=d\log g/d\log T$ necessarily tends to zero, and it no longer exposes the \emph{rate} at which the coupling approaches its fixed point. To quantify this approach we linearize the flow around $g_\star$ and extract the corresponding decay exponent.

In the Wishart-type model, the running normalization $\gamma^2(T)$ is fixed by the renormalization condition on the first moment
\begin{align}
  m_1=\frac{1}{\gamma^2(T)}\sum_{t=1}^T t^{-2\alpha}\stackrel{!}{=}m_1^\star\,,
\end{align}
so that, up to the fixed constant $m_1^\star$ factor
\begin{align}
  \gamma^2(T)\propto \sum_{t=1}^T t^{-2\alpha}\,.
\end{align}
For $\alpha>1/2$ the series converges and $\gamma^2(T)$ saturates as $T\to\infty$.

\subsection{RG fixed point approach}
We define the fixed-point normalization and coupling
\begin{align}
  \gamma_\star^2:=\lim_{T\to\infty}\gamma^2(T)\propto \zeta(2\alpha)\,,\quad
  g_\star:=\gamma_\star^{-2}\,,
\end{align}
and measure the approach to the fixed point via the deviation in $\gamma^2$
\begin{align}
  \delta(T):=\gamma_\star^2-\gamma^2(T)\,.
\end{align}
Using a tail estimate for the convergent series, we obtain the asymptotic scaling of the deviation
\begin{align}\label{eq:delta_scaling}
  \delta(T)\propto\sum_{t=T+1}^\infty t^{-2\alpha}\sim \int_T^\infty t'^{-2\alpha}\,dt'=\frac{T^{1-2\alpha}}{2\alpha-1}\,.
\end{align}
Hence, the deviation decays as a power law
\begin{align}
  \delta(T)\propto T^{-(2\alpha-1)}\,,
\end{align}
with exponent $2\alpha-1>0$.

\subsection{Linearized Beta function}
A natural linearized Beta function is the logarithmic derivative of the deviation:
\begin{align}\label{eq:beta_delta_def}
  \beta_\delta:=\frac{d\log\delta}{ds}=\frac{d\log\delta}{d\log T}\,.
\end{align}
Using Eq.~\eqref{eq:delta_scaling} we find
\begin{align}
  \beta_\delta\to 1-2\alpha<0\,,\quad \alpha>1/2\,.
\end{align}
Thus, in the saturating regime the flow is irrelevant in the RG sense: as the scale $T$ increases the deviation from the fixed point decays as a power law, and under decimation (reducing the scale) the deviation grows with exponent $2\alpha-1$.
Equivalently, we may linearize the coupling itself. Since $g(T)=\gamma^{-2}(T)$ and $\gamma^2(T)=\gamma_\star^2-\delta(T)$, we have
\begin{align}
  g(T)=\frac{1}{\gamma_\star^2-\delta(T)}=g_\star+\frac{\delta(T)}{\gamma_\star^4}+O\big(\delta(T)^2\big)\,,
\end{align}
so that $g(T)-g_\star\propto \delta(T)$ and therefore $g(T)-g_\star\propto T^{1-2\alpha}$ with the same exponent.

Thus, the linearized Beta function $\beta_\delta$ characterizes the approach to the fixed point through the deviation $\delta(T)$ (or, equivalently, $g(T)-g_\star$), and recovers the negative exponent $1-2\alpha$ in the saturating regime.

\section{Large-$N$ limit and the continuum fixed point}\label{app:large-N-limit-Wishart}
\begin{figure}[t]
\centering
\subfigure[Continuum solution and simulation.]{
\label{fig:9a}
\includegraphics[width=8.4cm]{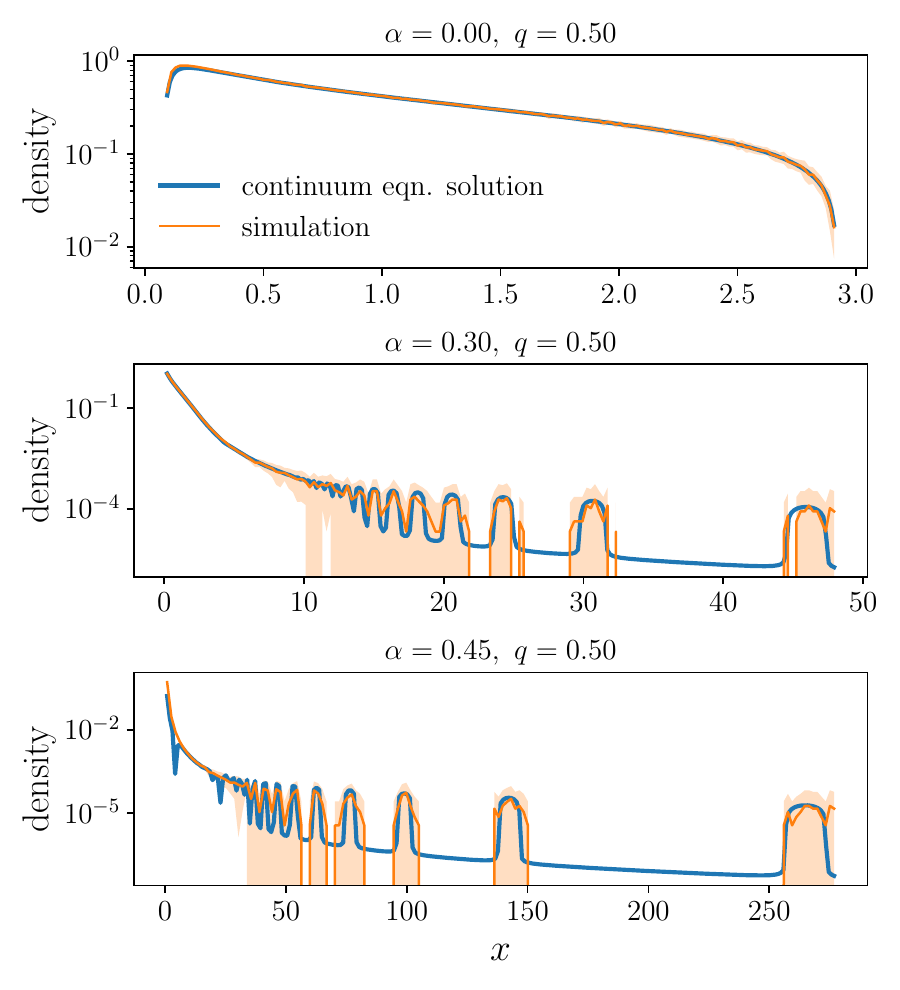}
}\\

\subfigure[Single-trial sampling of eigenvalue bands.]{
\label{fig:9b}
\includegraphics[width=8.4cm]{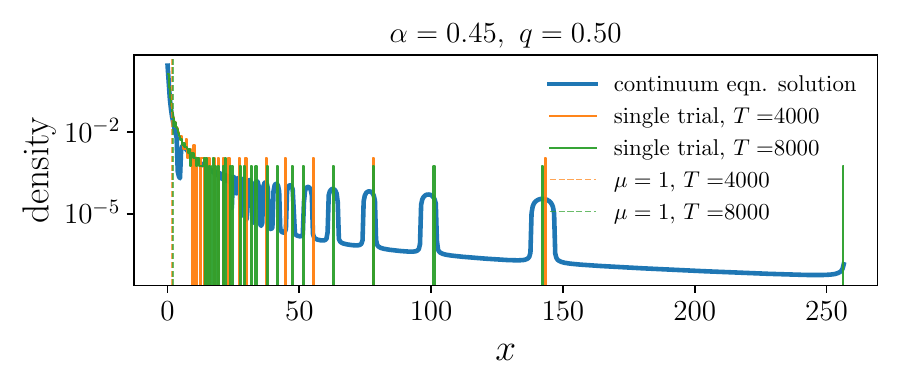}
}
\caption{{\bf Comparison of continuum equation solution and simulation.} (a) Plots show the continuum limit solution (blue) and the simulation with mean and one standard deviation band (orange) for $\alpha=0$, $0.3$, and $0.45$. Simulation curves are obtained with $N=4000$ and $40$ independent trials. (b) Continuum limit solution for $\alpha=0.45$ (blue) and single-trial simulations for $T=4000$ (orange) and $8000$ (green). Vertical lines show the sampling threshold beyond which single‑trial smoothing is sampling‑limited.
}\label{fig:continuum_solution_wishart}
\end{figure}
We take the large‑$N$ limit with fixed aspect ratio $q:=N/T$ and pass from row indices $t=1,\dots,T$ to the variable $y_t:=t/T\in(0,1]$ which becomes the continuous variable $y$ for large $T$. For the row-heterogeneity, $v_t=t^{-2\alpha}$, we have
\begin{align}\label{eq:v_y}
  v_t = (Ty_t)^{-2\alpha} = T^{-2\alpha}\,y_t^{-2\alpha}
\end{align}
and further define the scaled row weights
\begin{align}
  \tilde v_t:=\frac{T}{\gamma^2(T)}v_t=\frac{T^{1-2\alpha}}{\gamma^2(T)}\,y_t^{-2\alpha}\,.
\end{align}
In the continuum limit where $y_t\to y$, the analogue expression is
\begin{align}
    \tilde v_T(y):=\frac{T^{1-2\alpha}}{\gamma^2(T)}y^{-2\alpha}\,.
\end{align}
For $\alpha\neq 1/2$ the pointwise limit exists for fixed $y\in(0,1]$:
\begin{align}
    \tilde v(y) := \lim_{T\rightarrow\infty} \tilde v_T (y)\,.
\end{align}
Starting from the self-consistent Eqs.~\eqref{eq:wishart_finiteN_G}-\eqref{eq:wishart_finiteN_Sigma}, the sum giving the self-energy becomes an integral and the continuum closure takes the form
\begin{align}
  \overline G(z) &= \frac{1}{z-\Sigma(z)}\,,\\
  \Sigma(z) &= \int_0^1 \frac{\tilde v(y)}{1 - q\,\tilde v(y)\,\overline G(z)}\,dy\,.
  \label{eq:wishart_continuum_closure_correct}
\end{align}
with the physical branch chosen by $\Im\,\overline G(z)<0$. Next, we treat the different regimes of $\alpha$.

For $0\le\alpha<1/2$ we have $\gamma^2(T)\sim \frac{T^{1-2\alpha}}{1-2\alpha}$, and hence
\begin{align}
  \tilde v(y)=(1-2\alpha)\,y^{-2\alpha}\,.
\end{align}
As a sanity check, for $\alpha=0$ we have $\tilde v=1$ and
$\Sigma = 1/(1- q\,\overline G)$.
Then Eq.~\eqref{eq:wishart_continuum_closure_correct} reproduces the self-consistent equation for the Mar\v{c}enko--Pastur resolvent~\cite{potters2020first}.

In the large-$N$ continuum limit, the integral equation is derived at fixed $x=O(1)$ and for $y=t/T$ bounded away from $0$. For $\alpha>1/2$ one has $\tilde v(y)=0$ for any fixed $y\in(0,1]$, hence $\Sigma(z)=0$ and $\overline G(z)=1/z$ (the free resolvent) on the $O(1)$ spectral scale. The sector $t=O(1)$, or equivalently $y\sim 1/T$, forms a boundary layer that is not captured by the continuum limit and produces the outliers of size $O(N)$ that keep $m_1$ fixed.

Finally, the marginal case $\alpha=1/2$ requires keeping the logarithmic $T$-dependence. Since $\gamma^2(T)\sim\log T$, we have
\begin{align}
  \tilde v_t=\frac{1}{\log T}\,y_t^{-1}\,.
\end{align}
In particular, the continuum closure retains the lower cutoff $y_{\min}=1/T$ inherited from $t\ge1$ and reads
\begin{align}
  G_T(z) &= \frac{1}{z-\Sigma_T(z)}\,,\\
  \Sigma_T(z) &= \int_{1/T}^1\frac{\tilde v_T(y)}{1-q\,\tilde v_T(y)\,G_T(z)}\,dy\,,
  \quad
  \tilde v_T(y)=\frac{1}{\log T}\,y^{-1}\,.
\end{align}
Thus, $\alpha=1/2$ is marginal with logarithmic corrections. Although $\tilde v_T(y)\to0$ pointwise for fixed $y\in(0,1]$, the self-energy integral remains cutoff-sensitive and can contribute at $O(1)$ through $y_{\min}=1/T$ and the slow $1/\log T$ running. As a consequence, in the strict limit most eigenvalues accumulate near zero, while a vanishing fraction of eigenvalues remain away from zero on an $O(1)$ scale so that the mean level stays fixed.

In Fig.~\ref{fig:9a} we show a comparison of eigenvalue densities from simulated data with solutions of the large-$N$ continuum integral equation for $\alpha<1/2$. Overall, we find good agreement, but observe that the solutions of the integral equations tend to show some numerical instabilities leading to small local mismatches.
The simulated eigenvalue bands at large $x$ visible in the second and third subplots of Fig.~\ref{fig:9a} are produced by resonances of heavy rows with weights $v_t=t^{-2\alpha}$. Each band corresponds to a fixed row fraction $y=t/T$ and its $x$-position is $T$-independent (set by $1-q\,\tilde v(y)\,\overline G(z)\approx0$, cf. Eq.~\eqref{eq:wishart_continuum_closure_correct}), while the number of visible bands increases with $T$ as the $y$–grid refines. Each band carries a mass $O(1/T)$, and their Cauchy–smeared aggregate converges to the continuum bulk envelope (blue). In Fig.~\ref{fig:9b} we show that continuum eigenvalue bands are ``under‑filled'' by single-trial simulation data when $\mu(x):=N\,\eta\,\rho(x)\lesssim 1$. Increasing $\eta$, averaging over independent trials, or making $N$ very large fills the envelope without changing the bulk shape.


\providecommand{\noopsort}[1]{}\providecommand{\singleletter}[1]{#1}%

\end{document}